\documentclass{aa}  

\usepackage{booktabs} 
\usepackage{graphicx}
\usepackage{txfonts}
\usepackage{xspace}
\usepackage{orcidlink} 
\DeclareRobustCommand{\ion}[2]{\textrm{#1\,\textsc{#2}}}
\newcommand{\feii}{\ion{Fe}{ii}}
\newcommand{\ciii}{\ion{C}{III}]~$\lambda$1909\xspace}
\newcommand{\siiii}{\ion{Si}{III}]~$\lambda$1892\xspace}
\newcommand{\heii}{\ion{He}{II}~$\lambda$1640\xspace}

\newcommand{\mgii}{\ion{Mg}{II}~$\lambda$2800\xspace}
\newcommand{\oiii}{[\ion{O}{III}]~$\lambda$5007\xspace}

\newcommand{\oiiidd}{[\ion{O}{III}]~$\lambda\lambda$4959,5007\xspace}
\newcommand{\hbeta}{H$\beta$~$\lambda$4861\xspace}
\newcommand{\hbetashort}{H$\beta$\xspace}
\newcommand{\hgammafull}{H$\gamma$~$\lambda$4340\xspace}
\newcommand{\hgamma}{H$\gamma$\xspace}
\newcommand{\civ}{\ion{C}{IV}~$\lambda$1549\xspace}
\newcommand{\lya}{Ly$\alpha$~$\lambda$1216\xspace}

\newcommand{\pab}{Pa$\beta$~$\lambda$12818\xspace}

\newcommand{\nii}{[\ion{N}{II}]~$\lambda$6583\xspace}
\newcommand{\sii}{[\ion{S}{II}]~$\lambda\lambda$6716,6731\xspace}
\newcommand{\cii}{[\ion{C}{II}]~$\lambda$157.6\,$\mu$m\xspace}
\newcommand{\siv}{\ion{S}{IV}~$\lambda\lambda$1394,1403\xspace}
\newcommand{\halpha}{H$\alpha$~$\lambda$6563\xspace}
\newcommand{\oiiiaur}{[\ion{O}{III}]~$\lambda$4363\xspace}

\begin{document} 

    \titlerunning{JWST reveals kiloparsec-scale metal-free Balmer halo around a z=7.64 quasar}
\authorrunning{J. Wolf et al.}

   \title{Shedding the envelope: JWST reveals a kiloparsec-scale [O{\sc iii}]-weak Balmer shell around a z=7.64 quasar}

   \author{Julien Wolf
          \inst{1}\orcidlink{0000-0003-0643-7935},
           Eduardo Ba\~{n}ados \inst{1}\orcidlink{0000-0002-2931-7824},
           Xiaohui Fan
           \inst{2}\orcidlink{0000-0003-3310-0131},
           Antoine Dumont
           \inst{1}\orcidlink{0000-0003-0234-3376},
           James E. Davies
           \inst{1}\orcidlink{0000-0002-5079-9098},\\
           David S. N. Rupke   
           \inst{3,4}\orcidlink{0000-0002-1608-7564},
           Jinyi Yang
           \inst{5}\orcidlink{0000-0001-5287-4242},
           Weizhe Liu
           \inst{2}\orcidlink{0000-0003-3762-7344}
,
           Silvia Belladitta
           \inst{1}\orcidlink{0000-0003-4747-4484},
           Aaron Barth
           \inst{6}\orcidlink{0000-0002-3026-0562},
           Sarah Bosman \inst{1,7}\orcidlink{0000-0001-8582-7012},\\
           Tiago Costa\inst{8}\orcidlink{0000-0002-6748-2900},
            Frederick B. Davies
            \inst{1}\orcidlink{0000-0003-0821-3644},
            Roberto Decarli
           \inst{8}\orcidlink{0000-0002-2662-8803},
           Dominika {\v{D}}urov{\v{c}}{\'\i}kov{\'a}
           \inst{10}\orcidlink{0000-0001-8986-5235},
           Anna-Christina Eilers \inst{10}\orcidlink{0000-0003-2895-6218},\\
           Hyunsung D. Jun
           \inst{11,12}\orcidlink{0000-0003-1470-5901},
            Yichen Liu
           \inst{2}\orcidlink{0000-0003-4247-0169},
           Federica Loiacono
           \inst{9}\orcidlink{0000-0002-8858-6784},
           Alessandro Lupi
           \inst{13,14,8}\orcidlink{0000-0001-6106-7821},
           Chiara Mazzucchelli \inst{15}\orcidlink{0000-0002-5941-5214},
           Maria Pudoka
           \inst{2}\orcidlink{0000-0003-4924-5941},
           Sofia Rojas  \inst{16}\orcidlink{0000-0003-2349-9310},   
           Jan-Torge Schindler \inst{17}\orcidlink{0000-0002-4544-8242},
           Wei Leong Tee \inst{2}\orcidlink{0000-0003-0747-1780},
          Benny Trakhtenbrot
          \inst{18}\orcidlink{0000-0002-3683-7297}, \\
          Fabian Walter
          \inst{1}\orcidlink{0000-0003-4793-7880},
          Huanian Zhang \inst{19}\orcidlink{0000-0002-0123-9246}
          }

\institute{%
Max-Planck-Institut für Astronomie, Königstuhl 17, D-69117 Heidelberg, Germany
\and
Steward Observatory, University of Arizona, 933 N. Cherry Ave, Tucson, AZ 85719, USA
\and
Department of Physics, Rhodes College, 2000 North Parkway, Memphis, TN 38112, USA
\and
Zentrum für Astronomie der Universität Heidelberg, Astronomisches Rechen-Institut, Mönchhofstr 12-14, D-69120 Heidelberg, Germany
\and
Department of Astronomy, University of Michigan, 1085 S. University Ave., Ann Arbor, MI 48109, USA
\and
Department of Physics and Astronomy, University of California, Irvine, 4129 Frederick Reines Hall, Irvine, CA 92697-4575, USA
\and
Institute for Theoretical Physics, Heidelberg University, Philosophenweg 12, D-69120 Heidelberg, Germany
\and
School of Mathematics, Statistics and Physics, Newcastle University, Newcastle upon Tyne, NE1 7RU, UK
\and
INAF – Osservatorio di Astrofisica e Scienza dello Spazio di Bologna, Via Gobetti 93/3, I-40129 Bologna, Italy
\and
MIT Kavli Institute for Astrophysics and Space Research, 77 Massachusetts Avenue, Cambridge, MA 02139, USA
\and
Department of Physics, Northwestern College, 101 7th St SW, Orange City, IA 51041, USA
\and
School of Physics, Korea Institute for Advanced Study, 85 Hoegiro, Dongdaemun-gu, Seoul 02455, Republic of Korea
\and Como Lake Center for Astrophysics, DiSAT, Università degli Studi dell’Insubria, via Valleggio 11, I-22100, Como, Italy
\and INFN, Sezione Milano-Bicocca, P.za della Scienza 3, I-20126 Milano, Italy
\and Instituto de Estudios Astrofísicos, Facultad de Ingeniería y Ciencias, Universidad Diego Portales, Avenida Ejercito Libertador 441,
Santiago, Chile
\and Department of Physics and Astronomy, University of California, Los Angeles, 430 Portola Plaza, Los Angeles, CA 90095, USA
\and
Hamburg Observatory, Gojenbergsweg 112, D-21029 Hamburg, Germany
\and
School of Physics and Astronomy, Tel Aviv University, Tel Aviv 69978, Israel
\and Department of Astronomy, School of Physics, Huazhong University of
Science and Technology, Luoyu Road, Wuhan, 430074, Hubei, China%
}

\date{Received October 31, 2025; accepted XXXX YY, 202Z}

  \abstract{
   Luminous quasars at the redshift frontier $z>7$ serve as stringent probes of super-massive black hole (SMBH) formation and they are thought to undergo much of their growth obscured by dense gas and dust in their host galaxies. Fully characterizing the symbiotic evolution of SMBHs and hosts requires rest-frame optical observations that span spatial scales from the broad-line region (BLR) to the interstellar and circumgalactic medium (ISM/CGM). The James Webb Space Telescope (JWST) now provides the necessary spatially resolved spectroscopy to do so. But the physical conditions that regulate the interplay between SMBHs and their hosts at the highest redshifts, especially the nature of early feedback phases, remain unclear.   
   We present JWST/NIRSpec Integral Field Unit (IFU) observations of J0313$-$1806 at $z=7.64$, the most distant luminous quasar known. 
   From the restframe optical spectrum of the unresolved quasar, we derive a black hole mass of $M_\mathrm{BH}=(1.63 \pm 0.10)\times10^9 M_\odot$ based on \hbeta (\hbetashort) and an Eddington rate of $\lambda=L/L_\mathrm{Edd}=0.80\pm 0.05$, consistent with previous \mgii-based estimates. J0313--1806 exhibits no detectable [O III]$\lambda\lambda4959,5007$ emission on nuclear scales (3$\sigma$ upper limit equivalent width of \oiii $<1.42\, \AA$). 
Most remarkably, we detect an ionized gas shell extending out to $\sim 1.8$ kpc traced by \hbetashort emission that also lacks any significant [O III]$\lambda\lambda4959,5007$, with a $3\sigma$ upper limit on the [O III]$ \lambda$5007 to \hbetashort flux ratio of  $\log_{10} \left( F(\mathrm{[OIII]})/F(\mathrm{H}\beta)\right)=-1.15$. 
    Through photoionization modelling, we demonstrate that the extended emission is consistent with a thin, clumpy outflowing shell where [O III] is collisionally de-excited by dense gas. We interpret this structure as a fossil remnant of a recent blowout phase, providing evidence for episodic feedback cycles in one of the earliest quasars. 
    These findings suggest that dense ISM phases may play a crucial role in shaping the spectral properties of quasars accross cosmic time.}


   \maketitle
%
%-------------------------------------------------------------------

\section{Introduction}

In the luminous quasar phase, the powerful panchromatic radiation of active galactic nuclei (AGN) unambiguously traces the accretion of a supermassive black hole (SMBH) at the centre of galaxies. Through dedicated optical and near-infrared (NIR) surveys, quasars have been discovered at ever increasing redshifts, well within the first gigayear of the Universe \citep[$z>5.6$, e.g.,][]{fan01,reed15,wu15,banados16,banados22,jiang16,matsuoka16,matsuoka18,matsuoka18b,matsuoka19,matsuoka22,matsuoka25,wang17,gloudemans22,wolf24,ighina24,belladitta25}.  The SMBHs powering these distant quasars populate the extreme end of the black hole mass scale  \citep[$10^8-10^{10} \, M_\odot$, e.g.,][]{onoue19,yang21,farina22,mazzuchelli23} and thus provide stringent constraints on the formation and evolution of the very first massive black holes \citep{inayoshi20,fan23}.
Only a handful of quasars beyond $z>7$ have been identified to date \citep{mortlock11,banados18,yang20b,wang21}. The scarcity of confirmed sources at the quasar redshift frontier is generally attributed to two main factors:
\textit{(1)} the intrinsic steep decline in the space density of luminous quasars with increasing redshift, as captured by the quasar luminosity function at $z>6$ \citep[e.g.,][]{schindler23,matsuoka23lum}; and
\textit{(2)} the fact that quasars at $z\sim 7.1-7.5$ exhibit optical/NIR colours that are nearly indistinguishable from those of Galactic L- and T-dwarfs \citep[e.g.,][]{hewett06,lodieu07,mortlock09,fan23,banados25}.

The recent discovery of reionization-era blazars \citep{belladitta20,ighina24,wolf24,banados24,marcotulli25}, however, suggests a third possibility for the rarity of $z>7$ quasars: many of these objects may be systematically missed in restframe UV searches because they are undergoing obscured growth, enshrouded in dense gas and/or dust \citep[e.g.,][]{maiolino95,assef15,vito19b,ni20,lambrides20,gilli22}. The fraction of obscured quasars has been shown to increase with redshift \citep[e.g.,][]{buchner15,liu17,glikman18,lanzuisi18,vito18,iwasawa20,gilli22,peca23}. Indeed, from the \lya absorption profiles of quasars at $z>7$, \citet{davies19} inferred extremely low radiative efficiencies ($\leq 0.1\%$), which may indicate that these quasars remain obscured over nearly the entirety of their growth phases. They constrain the obscured fraction at $z>7$ to $>82\%$ at $95\%$ credibility.

This opens an exciting perspective, pointing to the existence of a vast, yet largely unexplored, quasar population at cosmic dawn. Conversely, the few archetypal dust-unobscured, broad-line quasars discovered at $z>7$ must have already cleared the line of sight to their nuclei and broad-line regions (BLR), perhaps through powerful feedback associated with the so-called “blowout phase” \citep[e.g.,][]{hopkins06,zakamska16,ishibashi18,lansbury20,vayner25}. Detecting observational signatures of such a recent blowout in the immediate environments of the highest-redshift quasars—for instance, fossil distributions of expelled dense gas or dust— would offer a robust anchor for embedding these extremely rare sources within a broader evolutionary sequence.

The James Webb Space Telescope (JWST, \citealt{gardner06})/Near-Infrared Camera (NIRCam, \citealt{rieke23}) has ushered in a new era by enabling, for the first time, detection of restframe ultra-violet (UV) and optical emission from the host galaxies of quasars at $z > 6$ \citep{ding22,ding23,stone23,yue24}. To overcome the stark contrast between the unresolved, dominant quasar light and the compact, faint stellar continuum emission at these wavelengths and redshifts, a detailed model of the instrument's point-spread function (PSF) is essential \citep[e.g.,][]{mechtley12}. With advanced imaging PSF subtraction techniques, it is now possible to directly measure the host galaxy's stellar mass from the extended continuum emission detected with JWST, and to compare it with the central SMBH mass to assess whether their growth was offset with respect to local scaling relations \citep[e.g.][]{magorrian98,ferrarese00,gebhardt00,haering04,kormendy13,reines15,habouzit21}. The JWST Near-Infrared Spectrograph Integral Field Unit (NIRSpec IFU, \citealt{boeker22}) enables integral field spectroscopy over tens of kiloparsecs around high-redshift quasars and AGN, providing simultaneous access to the host galaxy's stellar continuum, the quasar spectrum (e.g. \citealt{loiacono24}), and gas kinematics in the quasar vicinity—revealing quasar feedback \citep[e.g.][]{liu24} and large-scale interactions with companion galaxies \citep[e.g.,][]{marshall23,marshall24,decarli24,uebler24}.
Similarly to NIRCam, detailed PSF modelling techniques are required to disentangle the quasar from underlying gas emission \citep[e.g.,][]{wylezalek22,veilleux23,perna23,vayner24,marshall23,marshall24}.

At $z_\mathrm{[C\,II]}=7.6423$, the quasar J031343.84$-$180636.4 (hereafter J0313$-$1806) is currently the most distant system known in which the properties of a matured $>10^8 \, M_\odot$ SMBH, its host and direct environment can be studied in detail \citep{wang21}. The virial single-epoch estimate of the SMBH mass, $M_\mathrm{BH}$, based on the broad \mgii  emission line is $(1.6 \pm 0.4) \times 10^9 \, M_\odot$ and Atacama Large Millimeter Array (ALMA) observations indicate a high cold dust mass of $\sim 7 \times 10^7 \, M_\odot$. From its \cii emission, \citet{wang21} further derive a high star-formation rate $\mathrm{SFR}_\mathrm{[C\,II]}$ = 40-240 $M_\odot \mathrm{\, yr^{-1}}$. In its restframe UV spectrum, the object shows signatures of strong outflows through broad absorption lines in \civ, \siv and tentatively in \mgii (up to 0.19\textit{c}), as well as a significant blueshift of its \civ broad emission line ($\sim$3100 $\mathrm{km \, s^{-1}})$. The immediate environment of this quasar is therefore expected to show complex kinematics. 

We present a study aimed at resolving the kinematic structure of the extended, restframe optical ionized gas emission within $\sim$15 kpc of J0313$-$1806 in both spectral and spatial dimensions with the JWST/NIRSpec Integral Field Unit (IFU, Cycle 1 GO 1764). % We find an extended structure within $\sim 1$ kpc emitting  \hbetashort lines with additional absorption signatures. No \oiii emission is detected from this region. We interpret this as a clumpy, dense and optically gas envelope around, currently being expelled by strong outflows. We discuss how the properties of this shell potentially relate to the dense gas envelopes proposed by \citet{inayoshi25,kido25}. 
In Section \ref{sec:2Obs}, we describe the JWST/NIRSpec IFU observations of J0313--1806 and the associated data reduction. Section \ref{sec:3QuasarSpec} focuses on the nuclear quasar spectrum, from which we derive the central SMBH properties. In Section \ref{sec:4Extended}, we detail our PSF subtraction procedure and recover the extended gas emission in the quasar’s immediate environment. Finally, in Section \ref{sec:discussion}, we synthesize the results from the nuclear spectrum and the extended emission to place them in a broader physical context. We present our conclusions in Section \ref{sec:conclusions}.

Throughout this paper, we assume a concordance $\Lambda$CDM cosmology with parameters: $H_0=70 \, \mathrm{km\, s^{-1}}$, $\Omega_M=0.3$, $\Omega_\Lambda=0.7$. With this cosmology, one second of arc  corresponds to $4.96$ kpc in proper (physical) transverse distance at $z=7.64$. All equivalent widths are restframe measurements. All images are shown in the north-up, east-left convention.

\section{Observations and data reduction}
\label{sec:2Obs}
\subsection{JWST/NIRSpec IFU pointings}

We observed J0313$-$1806 and an associated PSF calibration star, TYC 5875-488-1, with JWST/NIRSpec IFU on Jan 21, 2023, as part of our JWST Cycle 1  proposal ID 1764. 
The total exposure times were 9.82 and 0.17 hours, respectively. 
The grating and filter combination G395M/F290LP covers wavelength range 2.87–5.10\,$\mu$m with a nominal resolving power of $R\sim 1000$.  The velocity resolution of our cube is thus roughly $300 \, \mathrm{km\,s^{-1}}$. The field of view of JWST/NIRSpec IFU is $3''\times 3''$ (corresponding to $\sim \, 15\, \mathrm{kpc} \times 15 \, \mathrm{kpc}$ at the redshift of the quasar), with a native pixel scale of $0.1''$.
The NIRSpec IFU PSF is undersampled by design. A \texttt{small} cyclic 8-point dithering pattern was chosen to improve spatial sampling. This observing strategy alleviates aliasing effects arising from the undersampled PSF \citep[so-called \textit{wiggles}, e.g.][]{perna23}. Each dither consists of 3 integrations of 20 groups each.  The readout pattern was set to NRSIRS2. In addition to the cyclic dithers, an additional {\it leakcal} exposure was taken to monitor and correct for  light leaks from the neighbouring micro-shutter-array (MSA) on JWST.

\subsection{Data reduction}

We reduced the data cubes with the standard three-stage JWST Science Calibration Pipeline \footnote{https://jwst-docs.stsci.edu/jwst-science-calibration-pipeline} version \texttt{1.15.0} (CRDS context: \texttt{jwst\_1322.pmap} ). The first stage, \texttt{calwebb\_detector1}, performs group-level corrections on uncalibrated data such as detector bias and dark subtractions, electronic noise, persistence and linearity corrections. It also detects cosmic rays (\texttt{jump}), converts counts to electrons, and fits a slope to the counts to calculate fluxes (\texttt{ramp\_fitting}). We inserted the additional processing step  \texttt{nsclean} between the cosmic-ray \texttt{jump} detection and \texttt{ramp\_fitting} steps\footnote{https://science.nasa.gov/mission/webb/for-scientists/\#NSClean}, which removes all correlated read noise from the frames. The output of the first calibration stage is a set of so-called rate images that form the input to the second stage \texttt{calwebb\_spec2}, which encompasses 2D spectroscopic processing steps such as flat-fielding, slitlet extraction, background subtraction, as well as wavelength and flux calibrations. The output of this stage is a set of 2D spectral images for each slitlet. The final stage, \texttt{calwebb\_spec3}, combines the individual spectra into a cube, ensuring astrometric coherence. An outlier rejection step removes residual artefacts by searching for pixels that display a sharp difference with respect to their neighbours in the spatial direction. 
The calibration stage includes three cube projection algorithms: 3D drizzle, weighting, and the estimated median signal method. We opted for 3D drizzle, which projects pixel flux from different dithers onto a common 3D grid, accounting for the relative pixel-to-pixel and pixel-to-voxel overlap. The major advantage of 3D drizzle is that it is particularly adapted to undersampled data by accessing sub-pixel scales, leveraging the input from multiple dithers. The cyclic dithering pattern allows us to reproject our data cube to a half-pixel scale of 0\farcs05 \footnote{https://jwst-docs.stsci.edu/jwst-near-infrared-spectrograph/nirspec-observing-strategies/nirspec-dithering-recommended-strategies\#gsc.tab=0}, thus improving spatial sampling. At the redshift of the quasar, $z=7.6423$, this corresponds to a projected physical scale of $\sim 0.2$ kpc/pixel. However, the code assumes a rigid one-to-one mapping of the illuminated chip x- and y-axis and the dispersion and spatial dimensions of the instrument. This introduces undersampling artefacts that were addressed in the additional pre-processing step detailed in Appendix \ref{sec:appendix_wicked}.

The red continuum imprinted by the zodiacal and stray light background was accounted for by sigma-clipping the reduced cubes in flux, masking three sigma outliers. The unmasked regions are treated as field spaxels. We extract the median 1D continuum from these spaxels and subtracted it from the cubes.

\section{Physical properties of the black hole}
\label{sec:3QuasarSpec}
\subsection{Spectral fitting}
\label{sec:spec_fit}
\begin{figure*}
    \centering
    \includegraphics[width=1.0\linewidth]{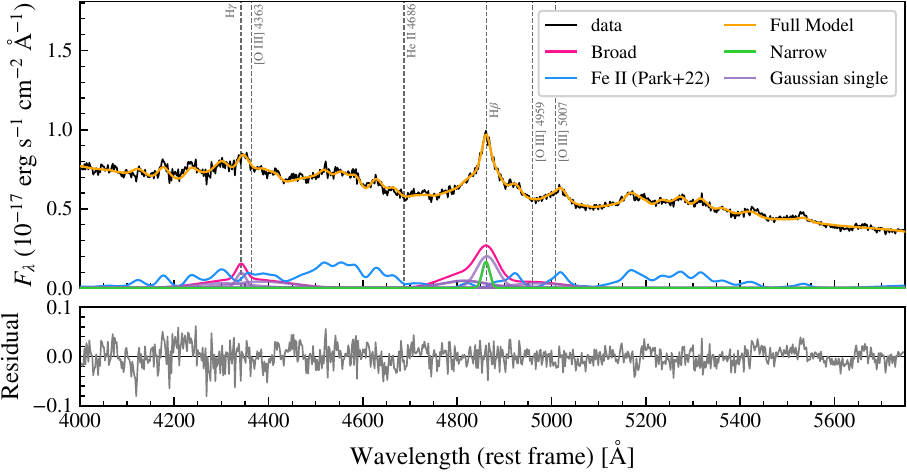}
    \caption{ Extracted quasar spectrum within a 0\farcs35 aperture centred on the spaxel with the highest integrated flux. The observed spectrum, shown in black, is well fit by the \texttt{PyQSOFit} model, shown in orange. Broad \hbetashort and \hgamma are modelled using three and two broad Gaussian components (pink lines, individual Gaussians are shown in purple) with FWHM$>1200$ km/s, and one Gaussian component with FWHM$<1200$ km/s to trace narrow components (green line).
    The \oiiidd emission lines are modelled with one broad and one narrow component. No narrow \oiiidd components are detected, while a strong contribution from Fe{\sc ii} emission (blue line) is evident \citep{park22}. The lower panel shows fit residuals.}
    \label{fig:quasar_spectrum}
\end{figure*}

To determine the properties of the SMBH powering J0313$-$1806, we apply the same methodology as \citet{liu24}.
We extracted the quasar spectrum by identifying the spaxel with the highest integrated flux, which was defined as the location of the unresolved quasar emission (pixel (49,46), 0-indexed) within an aperture of radius $0\farcs35$ .

Following \citet{loiacono24}, we determined the optimal extraction radius for the quasar spectrum by testing values that minimize background and extended emission contamination. Spectra of the PSF star TYC 5875-488-1 were extracted at radii from 3 to 14 pixels (0\farcs15 to 0\farcs70) in 0\farcs05 steps.
We calculate the wavelength-dependent encircled-flux fraction by dividing these extracted spectra by the spectrum extracted in a large reference aperture of 1\farcs 0 (20 pixels). For each radius, the chromatic flux loss (CFL) was defined as
\[
\mathrm{CFL}(R,\lambda) = \frac{f_{\mathrm{blue}} - f_{\mathrm{red}}}{f_{\mathrm{blue}}} \times 100\%,
\]
where \(f_{\mathrm{blue}}\) and \(f_{\mathrm{red}}\) are the median enclosed--flux fractions
measured in the blue (2.9--3.5\,$\mu\mathrm{m}$) and red (4.5--5.1\,$\mu\mathrm{m}$) bands, respectively. Wavelength--dependent flux fractions for all aperture radii are shown in Appendix~\ref{sec:appendix_extract}. We adopt an extraction radius of $0\farcs35$ \citep[see also][]{marshall23}, which captures the majority of the PSF flux while keeping the chromatic flux loss below the percent level and simultaneously limiting background and extended emission contamination. We note that this radius corresponds to $\sim$1.4 kpc at the redshift of the quasar and thus completely encompasses the BLR and the classical narrow-line region (NLR, e.g. \citealt{baskin05}). We model the wavelength-dependent correction by fitting a smooth spline to the 0\farcs 35 encircled fraction array and apply the correction to the quasar spectrum.

The resulting spectrum is presented in Fig.~\ref{fig:quasar_spectrum}. Notably, the spectrum reveals relatively weak \oiiidd\ emission lines and a strong Fe\,{\sc ii} pseudo-continuum, features that are also observed in the $z=7.5$ quasar J100758.26+211529.2 \citep{liu24}. These characteristics are typical of rapidly accreting quasars (see discussion in Sec.~\ref{sec:discussion}). 

To analyze the spectrum, we used the public software \texttt{PyQSOFit} \citep{guo18,shen19,ren24}. We selected featureless spectral regions ($4000-4200 \AA$, $4500-4700 \AA$,  and $5050-5750 \AA$), to fit the quasar continuum. Our setup uses a power-law, a restframe optical Fe{\sc ii} template \citep{park22} covering the \hbeta (\hbetashort) region and an additional smooth second-order polynomial correction term to adjust the tilt and curvature of the fit. We fixed the redshift of the quasar at $z=7.6423$ (as measured from \cii by \citealt{wang21}).
After subtracting the best-fit pseudo-continuum, we fitted the H$\beta$ emission line with three  broad Gaussians (FWHM $>1200\, \mathrm{km\,s^{-1}}$), as well as one narrow Gaussian (FWHM $<1200\, \mathrm{km\,s^{-1}}$). 
The Akaike (AIC) and Bayesian (BIC) information criteria both indicate that a three-component broad \hbetashort model is ideal. Adding additional components does not significantly improve the $\chi ^2$, while a two-component model underfits the line profile.   We fit  \hgammafull (\hgamma) with two broad and one narrow component.
The [OIII]$\lambda$4959 and [OIII]$\lambda$5007 lines are fit with one broad and one narrow Gaussian each. In all fits, the flux ratios of the \oiiidd doublet were constrained to their theoretical values, 
$F_{5007}/F_{4959} = 2.98$ (e.g., \citealt{storey00}). In our fit, the narrow core of the \oiiidd\ lines collapsed to zero flux. Very weak wings of the \oiiidd\ lines are detected (EW([OIII]$\lambda$5007$\approx 0.02 \, \AA$), however this is sensitive to the continuum placement and to the adopted decomposition of the \hbetashort base. We thus treat the \oiiidd as not significantly detected and quote a 3$\sigma$ upper limit on the equivalent width of [OIII]$\lambda$5007 measured from the local continuum noise, assuming a Gaussian line profile with a fixed FWHM of 19 $\AA$: EW([OIII]$\lambda$5007)<$1.42$ $\AA$ (based on the width of the detected narrow H$\beta$ component). The broad-line Balmer decrement shows no evidence for BLR reddening.
 The resulting best-fit spectrum, along with the fits for the broad emission lines and the Fe{\sc ii} model components, is presented in Fig.~\ref{fig:quasar_spectrum}.

\subsection{Black hole mass and Eddington rate}
\label{sec:qso}
Our best-fit broad Gaussian H$\beta$ model (i.e. excluding the narrow component) has an FWHM of $4156 \pm 68 \mathrm{km \, s^{-1}}$, which we interpret as virial broadening in the BLR \citep[e.g., ][]{peterson06}. The instrumental broadening of the NIRSpec G395M/F290LP configuration is negligible compared to the measured broad H$\beta$ width. Its error is dominated by the statistical uncertainty; we therefore do not apply an instrumental broadening correction. We measured the best-fit continuum monochromatic luminosity at restframe 5100\,\AA.  
To account for systematics, we added a conservative 10\% flux calibration uncertainty in quadrature to the small statistical error\footnote{We apply this 10\% uncertainty floor to all subsequent flux and equivalent width measurements.}, obtaining 
$L_{5100} = (1.77 \pm 0.18) \times 10^{46}\,\mathrm{erg\,s^{-1}}$.  
Together with the measured H$\beta$ line width, this allows us to estimate the black hole mass $M_{\mathrm{BH}}$ of the central SMBH using single-epoch calibrations, such as \citet[e.g.,][]{greene05,Vestergaard06,shen24}, following  \citet{Kaspi00}. We use the parametrisation of \citet{shen24}:
\begin{equation}
    \log \left( \frac{M_{\rm BH}}{M_\odot} \right) =
    0.85
    + 0.50 \,\log \!\left( \frac{L_{5100}}{10^{44}\,\mathrm{erg\,s^{-1}}} \right)
    + 2.0 \,\log \!\left( \frac{\mathrm{FWHM}_{\rm H\beta}}{\mathrm{km\,s^{-1}}} \right),
\end{equation}
with an intrinsic scatter of $\sigma_{\rm int} = 0.45 \pm 0.04$ dex. 

We obtain an H$\beta$ single-epoch mass of 
$M_{\rm BH,H\beta} = (1.63 \pm 0.10) \times 10^{9}\,M_\odot$, 
and 
$M_{\rm BH,H\beta} = (1.63^{+2.97}_{-1.05}) \times 10^{9}\,M_\odot$ 
when accounting for the intrinsic scatter.
We use the measurements of \mgii and $L_{3000}$ reported by \citet{wang21} to 
re-visit the black hole mass estimate using the \mgii single-epoch calibration of 
\citet{shen24}, yielding 
$M_{\rm BH,MgII} = (3.18 \pm 1.06) \times 10^{9}\,M_\odot$, 
or 
$M_{\rm BH,MgII} = (3.18^{+6.26}_{-2.11}) \times 10^{9}\,M_\odot$ 
when including intrinsic scatter. 
Considering measurement uncertainties alone, the two estimates differ at the 
$\sim1.8\sigma$ level. When the intrinsic scatter of the calibrations is included, the results are fully consistent. We note that using the \citet{shen24} calibration for the \mgii-based black hole mass yields values that are higher by a factor of $\sim2$ compared to those obtained with the \citet{vestergaard09} calibration adopted by \citet{wang21} and \citet{yang21}; the two estimates are consistent only when accounting for the intrinsic scatter of the relations.

We calculate the Eddington rate $\lambda_\mathrm{Edd}$, defined as the ratio of bolometric luminosity $L_{\mathrm{bol}}$ and the Eddington luminosity $L_{\mathrm{Edd}} = \frac{4 \pi G M_{\mathrm{BH}} m_p c}{\sigma_T}$, where $m_p$ is the proton mass, $c$ is the speed of light, and $\sigma_T$ is the Thomson scattering cross-section. We derive the bolometric luminosity using the bolometric correction factor from \citet{richards06}, applying $L_{\mathrm{bol}} = 9.26 \times L_{5100}$, which yields $(1.64 \pm 0.16) \times 10^{47}~\mathrm{erg\,s^{-1}}$. From our derived $M_\mathrm{BH}$ estimate 
we calculate an Eddington ratio of $\lambda_{\mathrm{Edd}} = 0.80 \pm 0.05$ accounting only for measurement uncertainties.
These findings corroborate the results of \citet{wang21}, indicating that quasar J0313---1806 harbours a relatively mature SMBH accreting at a high Eddington rate. The results of the spectral fit are summarized in Table~\ref{tab:spec}.

\begin{table}[]
\begin{tabular}{@{}lll@{}}
\toprule
Quantity                     & Units                   & Value                           \\ \midrule
FWHM$_\mathrm{H\beta}^{(1)}$ & $\mathrm{km\, s^{-1}}$  & $4156 \pm 68$                   \\
$L_{5100}^{(2)}$             & $\mathrm{erg\, s^{-1}}$ & $(1.77 \pm 0.18)\times 10^{46}$ \\
EW$(\mathrm{H\beta})^{(3)}$  & $\AA$                   & $53.18\pm 0.31$                 \\
EW$(\mathrm{[OIII]})^{(4)}$  & $\AA$                   & $<1.42$                \\
EW$(\mathrm{FeII})^{(5)}$     & $\AA$                   & $38.46\pm 0.02$                 \\
F$(\mathrm{[OIII]})^{(6)}$  & $\mathrm{\, erg \, s^{-1} cm^{-2}}$                   & $<2.39 \times 10^{-18} $                \\
$M_\mathrm{BH,H\beta}^{(7)}$ & $M_\odot$                & $(1.63 \pm 0.10)\times 10^9$     \\
$L_\mathrm{bol}^{(8)}$              & $\mathrm{erg\, s^{-1}}$ & $(1.64\pm 0.16)\times 10^{47}$ \\
$L_\mathrm{bol}/L_\mathrm{Edd}^{(9)}$            & --                      & $0.80 \pm 0.05$                  \\ \bottomrule
\end{tabular}
\caption{Results of spectral fitting for J0313--1806. (1) Full-width-at-half-maxium of the broad \hbetashort line (2) Monochromatic luminosity at 5100 $\AA$ (3) Equivalent width of the broad \hbetashort line (4) $3\sigma$ upper limit on the equivalent width of \oiii (5) Equivalent width of the \feii\ blend at 4434–4684$\AA$ 6) $3\sigma$ upper limit on the flux of \oiii (7) Black hole mass derived based on \citet{shen24} (8) Bolometric luminosity based on \citet{richards06} (9) Eddington rate}
\label{tab:spec}

\end{table}

\section{Extended nebular emission}
\label{sec:4Extended}

\subsection{Point-spread-function modelling}
\label{sec:psf_model}

To identify any potential extended gas components linked to the host galaxy, outflows, or field sources in the cube, we must first model and subtract the unresolved light from the significantly more luminous central quasar.  
Here, we create an empirical model of the PSF based on the observations of the associated PSF star, TYC 5875-488-1. To precisely align the quasar and PSF star images, we first compute the two-dimensional cross-correlation between the quasar and PSF star white-light images obtained by integrating the cubes over the wavelength range 3.5--5.1\,$\mu$m. This cross-correlation evaluates the similarity between the images at different pixel offsets.  

 We apply a PSF alignment shift between the brightest spaxels of the collapsed cubes, determining a sub-pixel shift in Fourier space \citep[e.g.,][]{stone01,foroosh02}.  This technique involves computing the Fourier transform of each wavelength slice, applying the calculated phase shift for sub-pixel precision, and then performing an inverse Fourier transform on the shifted data. We obtain \texttt{x-shift} = 1.48 pixels and \texttt{y-shift} $= -2.34$ pixels. The aligned PSF star cube is computed as:

\begin{equation}
    \tilde{I}_{\text{PSF}}(x, y) = \mathcal{F}^{-1} \left\{ \mathcal{F}[I_{\text{PSF}}(x, y)] \cdot e^{-2\pi i (x_{\text{shift}}, y_{\text{shift}})} \right\}  .
\end{equation}

\begin{figure*}
    \centering    \includegraphics[width=0.95\linewidth]{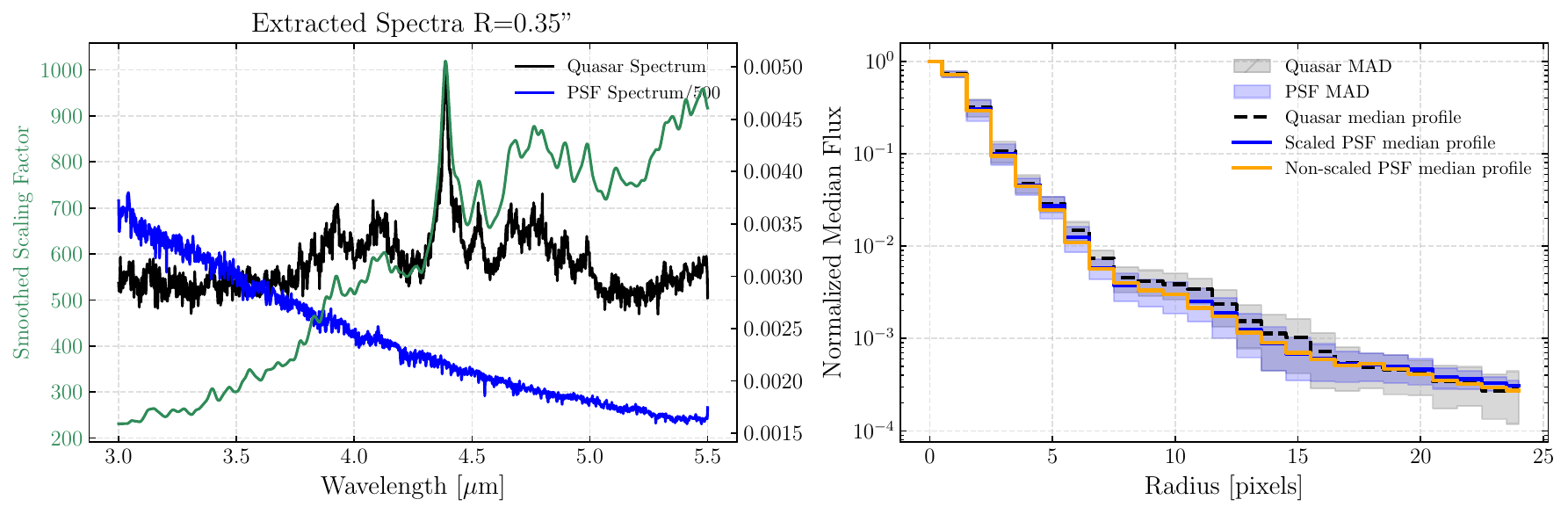}
    \caption{ PSF model. 
\textit{Left:} Spectral comparison between the quasar and PSF star used to derive the PSF scaling factor. The extracted 1D spectra from the quasar (black) and PSF star (blue) were obtained using a circular aperture with radius 0\farcs 35. The PSF spectrum is scaled down by a factor of 500 for visual comparison. While the PSF star shows a smooth stellar continuum, the quasar exhibits a prominent broad \hbetashort emission line and strong Fe\,{\sc ii} features. The wavelength-dependent scaling factor $S(\lambda)$ computed as the ratio between the quasar and PSF spectra is shown in green. The scaling factor is smoothed with a Gaussian kernel of width $\sigma = 5$ pix before being applied slice by slice to the aligned PSF cube. \textit{Right:} Median radial profiles extracted from the quasar cube (black dashed), the raw PSF star cube (orange) and the PSF star cube, corrected with the wavelength-dependent scaling factor (blue). The radial profiles are measured from the brightest pixel in white light of each cube. The colour-shaded areas correspond to the median absolute deviations of the quasar radial profile (grey) and the scaled PSF star cube (blue).}
    \label{fig:psfsub}
\end{figure*}

To account for the spectral energy distribution (SED) differences between the quasar and the PSF star, we extract integrated spectra using a circular aperture (radius = 7 pixels, i.e. 0$\farcs$35). %The size of this aperture is chosen to minimize the root-mean-square (RMS) of the resiudal map and ensures limited chromatic flux loss as discussed in Section \ref{sec:3QuasarSpec}. 
The aperture scale is determined based on the analysis shown in Appendix \ref{sec:appendix_extract}. At the chosen radius, the enclosed flux fraction is already within $\approx$1--2\% of that measured with larger apertures, and its wavelength dependence is smooth and consistent with the expected chromatic broadening of the JWST PSF. This indicates that the extracted flux is dominated by unresolved quasar light, with no evidence for a significant contribution from extended host emission.

The raw scaling factor per wavelength is given by $S(\lambda) = F_\mathrm{quasar}(\lambda)/F_\mathrm{PSF}(\lambda)$. However, small-scale fluctuations in $S(\lambda)$ can introduce systematic artifacts in the residual spectrum. To mitigate this, we apply a Gaussian smoothing filter with $\sigma$ = 5 pix along the spectral axis. The spectra and the smoothed wavelength-dependent scale factor are shown in the left panel of Fig. \ref{fig:psfsub}. The figure in the right panel shows that the PSF model, i.e. the smoothed and scaled PSF star, accurately traces the flux profile of the quasar. 

\begin{figure*}[t]  
    \centering
    \includegraphics[width=0.9\textwidth]{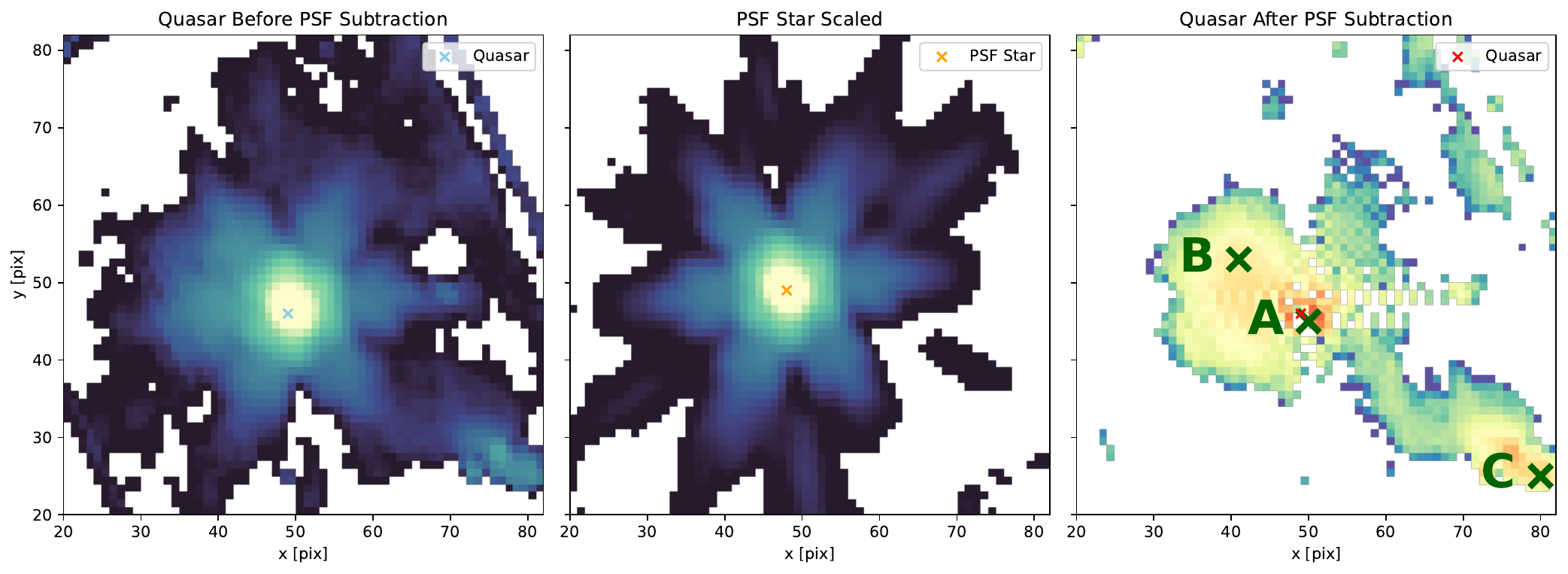}  
\caption{
PSF subtraction procedure for J0313--1806. \textit{ Left:} Integrated white-light image of the reduced cube centered on J0313--1806. The blue cross marks the position of the unresolved quasar. \textit{Centre:} Integrated and scaled cube of the PSF calibration star TYC 5875-488-1, aligned and scaled to match the quasar emission. The orange cross marks the position of the unresolved star. \textit{Right:} Residual white-light image of J0313--1806 after PSF subtraction, revealing extended emission components after removal of the unresolved quasar core. We highlight three regions of interest: A) A bright core near the quasar, B) a near-elliptical diffuse region just north-east of the quasar and C) a distinct foreground elliptical source to the south-west of the quasar. This work focuses on the central region A). For all maps, pixels with negative integrated flux were masked.
}
    \label{fig:wide_figure}
\end{figure*}

Once the smoothed scale factors are computed, we apply them to each wavelength slice of the PSF star cube, scaling it to match the quasar SED:
\begin{equation}
    I_{\text{PSF, scaled}}(x, y, \lambda) = S_{\text{smooth}}(\lambda) \times \tilde{I}_{\text{PSF}}(x, y, \lambda),
\end{equation}
where \( S_{\text{smooth}}(\lambda) \) denotes the smoothed wavelength-dependent scale factor, and \( \tilde{I}_{\text{PSF}}(x, y, \lambda) \) is the subpixel-aligned PSF cube.

The final PSF-subtracted quasar cube is then obtained by subtracting the scaled PSF cube from the observed quasar data:
\begin{equation}
    I_{\text{residual}}(x, y, \lambda) = I_{\text{quasar}}(x, y, \lambda) - I_{\text{PSF, scaled}}(x, y, \lambda).
\end{equation}

The resulting residual cube \( I_{\text{residual}}(x, y, \lambda) \) reveals the underlying spatially extended emission that is not associated with the unresolved quasar core. This emission can originate from narrow-line regions, the host galaxy, outflows, satellite companions or chance-aligned field fore- or background sources. A visual overview of the PSF subtraction process is presented in Fig.~\ref{fig:wide_figure}. 

Despite the careful modelling and subtraction procedure, several sources of systematic uncertainty remain. First, the SEDs of the quasar and the PSF star differ intrinsically, introducing wavelength-dependent mismatches that are not fully corrected by the smooth scaling procedure. Second, subpixel alignment errors, particularly in the presence of asymmetric PSF wings or optical distortions, can leave residual artifacts in the final cube. We have assessed the impact of these effects through a series of robustness tests. Varying the degree of spectral smoothing applied to the scaling factor $S(\lambda)$ results in negligible changes to the residual maps and does not affect the detection of extended emission. In addition, perturbing the relative alignment between the quasar and PSF star cubes at the level of the measured alignment uncertainty produces only minor changes in the residual root-mean-square (RMS). Residual artifacts associated with these systematics are confined to the central PSF core and do not generate spatially extended or spectrally coherent features. We therefore conclude that these systematic uncertainties do not affect the results or interpretations presented below.

In the collapsed PSF-subtracted cube, we identify three regions of interest. A luminous core close to the quasar position at pixel (50,45) (see green cross close to A); the object of the present paper. We further identify an extended quasi-elliptical, nebulous region to the north-east (B) of the quasar. The origin of this emission is further discussed in Appendix \ref{sec:appendix_cont}. We also find a distinct source to the south-west (C) is spatially co-incident with a foreground galaxy detected in JWST/Near Infrared Camera (NIRCam) and JWST/Mid-Infrared Instrument (MIRI)  imaging from the same Cycle 1 programme (\#1764). The extracted source has photometry consistent with that of a galaxy at $z\sim 2.25$ \citep[][and private communication]{pudoka25}. A clear detection in the F090W filter consolidates the lower redshift nature of this source. A spectrum extracted at this location displays a strong emission line, fixing the galaxy to $z=2.30$, through its \pab line. 
In Appendix \ref{sec:appendix_specs}, we present spectra extracted within 0\farcs35 of the emission regions A, B and C. The alternating horizontal pattern near region A is a PSF subtraction artifact caused by residual structured PSF wings (including diffraction features) and IFU sampling.

\subsection{Balmer kinematic moment maps}
\label{sec:moment_maps}

\begin{figure*}
    \centering
    \includegraphics[width=0.95\linewidth]{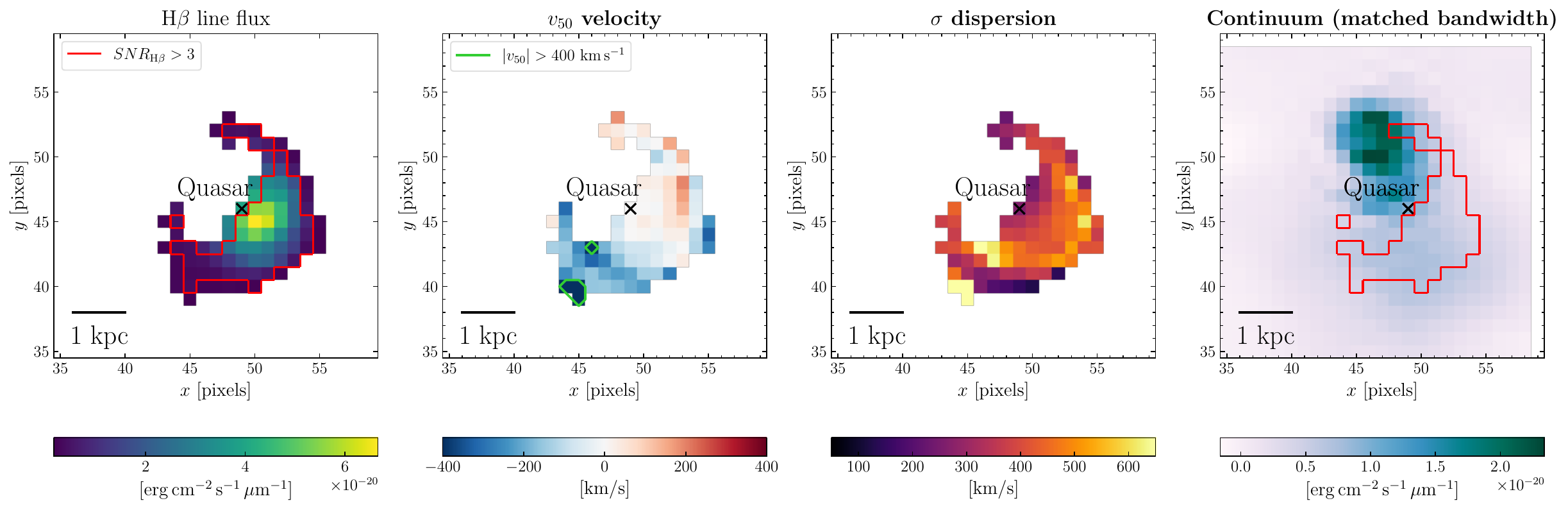}
    \caption{Kinematic moment maps of \hbetashort around J0313-1806 and bandwidth-matched continuum. Gaussian spatial smoothing over 2.5 pixel kernel was applied (at 0\farcs05/pix scale). \textit{First panel}: Flux map of $3\sigma$ detected \hbetashort (57 spaxels). These are highlighted by the red contour. We define this distribution as \hbetashort shell. Adjunct spaxels in 8-connectivity to these detection spaxels with at least S/N$_\mathrm{H\beta}> 1.5$ are also shown. The flux in the line peaks close to the quasar location, consistent with a central ionizing source. \textit{Second panel}: Velocity offset map as traced by the the $v_{50}$ parameter. A clear velocity gradient is observed along the shell, indicating expansion or rotation of the structure. Spaxels with extreme velocities ($\mid v_{50} \mid >400 \, \mathrm{km\, s^{-1}}$) are marked with green contours.  \textit{Third panel}: Velocity dispersion, $\sigma$. Patches of large dispersions (>600 $\mathrm{km/s}$) are observed at the edges and to the south-east of the structure, potentially indicating highly turbulent gas motion. \textit{Fourth panel}: Continuum map constructed by integrating the fitted continuum model over a line-free spectral window with bandwidth matched to that of \hbetashort. While elevated continuum emission overlaps spatially  the with the \hbetashort shell, the absence of brightened nuclear morphology in the map following the \hbetashort flux confirms that the observed \hbetashort structure is not driven by background variations or continuum residuals. The continuum map shows a clear flux peak offset to the north from the quasar position. This component does not trace the shell-like \hbetashort morphology and lacks any associated kinematic structure. We therefore interpret it as artifact from imperfect fitting in imperfectly PSF-subtracted spaxels, rather than extended stellar or nebular continuum emission.}
    \label{fig:moment_maps}
\end{figure*}

After PSF subtraction, we fitted the data cube with a combination of host continuum and emission line models using the cube analysis software package \texttt{q3dfit}\footnote{\url{https://github.com/Q3D/q3dfit}} \citep{rupke14,rupke21}, operating in \textit{fitpoly} mode. We apply Gaussian smoothing with a 2.5 pixel ($\sim 0.5$ kpc) kernel along the spatial dimensions to suppress pixel-to-pixel noise fluctuations and enhance the detectability of low–surface-brightness emission, while also mitigating aliasing effects.
 The continuum was modeled with a third-order polynomial, while emission lines were fitted with single Gaussian components. Initially, we restricted the line fitting to [O\,{\sc iii}]$\lambda$5007 and the Balmer lines H$\beta$ and H$\gamma$, with the Balmer lines kinematically tied. We stress that [O\,{\sc iii}]$\lambda$5007 and the Balmer lines are fitted individually, allowing the code to trace kinematically distinct regions.
Prior to fitting, the model spectra were convolved to match the spectral resolution of the G395M grating. We allow the code to fit lines up to $\sigma = 2,500\,\mathrm{km\,s^{-1}}$ (FWHM$\approx$5,900 km/s), a permissive upper bound for extended ionized gas around a quasar. We obtain zeroth, first and second order kinematic moments using \texttt{q3dfit}. The flux map are obtained from the single-Gaussian line fits.  No spaxel with \oiiidd (S/N>3 in flux) is detected. While we cannot exclude that this is an artifact of the PSF subtraction procedure, we treat \oiiidd\ as non-detected. For \hbetashort we apply a signal-to-noise threshold S/N$_\mathrm{H\beta}> 3$ on the 0-th moment maps. Adjunct spaxels in 8-connectivity to these detected spaxels with at least S/N$_\mathrm{H\beta}> 1.5$ are also conserved for plotting purposes only. The momemt maps are displayed in Fig. \ref{fig:moment_maps}. 
Pixels not fulfilling this criterion are masked. The
\hbetashort emission region forms an envelope or shell-like structure centred on the quasar. 57 spaxels have robustly detected \hbetashort with
S/N$_{H\beta}> 3$ and extend over a region of radius of $\sim  1.79$ kpc of the quasar position. We formally refer to this structure as the \hbetashort shell. We further identify 46 spaxels with $1.5 < \mathrm{S/N}_{\mathrm{H}\beta} < 3$, which are in 8-connectivity to neighboring spaxels in which  H$\beta$ is $3\sigma$ detected. While kinematically tied to \hbetashort, \hgamma is only
detected in 11/57 of the S/N$_\mathrm{H\beta}> 3$  spaxels at an integrated S/N$_\mathrm{H\gamma}> 2$.
The average projected distance of these spaxels to the quasar is 1.06 kpc. The velocity offset is traced by the $v_{50}$ parameter, i.e., the shift in velocity space at 50\% of the line intensity. We note that the detection of parts of the \hbetashort shell is also warranted with less aggressive spatial smoothing, e.g. a kernel of 1 pixel.

We first observe that the total fitted line flux map of \hbetashort shows an arc-like shape with a flux maximum in the center of the shell at the pixel (50,45), near the unresolved quasar position (49,46), supporting the presence of a central ionizing source, i.e. the actual AGN or a nuclear central star cluster. The slight offset between the H$\beta$-flux peak and the quasar position is within the FWHM of the JWST PSF and is attributable to residual PSF subtraction effects.
In Fig. \ref{fig:moment_maps}, the $v_{50}$ map displays a clear velocity dipole from blueshifts to redshifts south-east to north-west. We measure a median blueshifted ($v_{50}<0 \, \mathrm{km\, s^{-1}}$) velocity  $v_\mathrm{50,blue}= -113^{+17}_{-27} \, \mathrm{km s^{-1}}$ and a median redshifted ($v_{50}>0 \, \mathrm{km\, s^{-1}}$) velocity $v_\mathrm{50,red}= 32^{+4}_{-8} \, \mathrm{km \, s^{-1}}$. The gradient of the velocity field appears centred on the quasar. The offset values are typical of AGN outflows \citep[e.g.,][]{liu24}.The symmetric, ordered nature of the velocity field around the quasar indicates a coherent large-scale kinematic pattern, consistent with a tilted or partial expanding shell of ionized gas \citep[e.g.,][]{ishibashi21}, although other complex organized motions (e.g. rotation \citealt[][]{ishikawa25} or a biconical flow) cannot be excluded. The apparent arc-like geometry of the nebula might also be due to PSF-over-subtraction close to the quasar, effectively depleting signal in \hbetashort. For simplicity we will assume a spherical shell-like geometry (i.e. a complete sphere around the quasar) while acknowledging that this might not fully capture the real structure of the ionized gas. 
Finally, the dispersion map shows strong velocity dispersions, with a median dispersion of $\sigma = (415 \pm 51) \, \mathrm{km\, s^{-1}}$. Such dispersions are typically associated with AGN outflows and turbulent gas motion \citep[e.g.,][]{wylezalek22}. Spaxels corresponding to extreme velocity offsets in the $v_{50}$ map, i.e. $\mid v_{50} \mid >400 \, \mathrm{km\, s^{-1}}$ shown with green contours, appear to also correspond to the highest dispersions $\sigma > 600\, \mathrm{km\, s^{-1}}$. We note that we cannot fully exclude the possibility, that the observed \hbetashort distribution is directly tracing the host galaxy of J0313$-$1806 or a merging companion. Deep restframe UV/optical imaging with JWST/NIRCam will be required in the future to investigate the presence of a possible co-spatial stellar continuum. In Fig. \ref{fig:moment_maps} the control continuum map is constructed by integrating the continuum model over a line-free spectral window with bandwidth matched to that of \hbetashort (three times the median velocity $\sigma$ of \hbetashort), does not reproduce the peaked structure seen in the line emission, indicating that the observed structure is not driven by continuum or background systematics.

\begin{figure*}
    \centering
    \includegraphics[width=\linewidth]{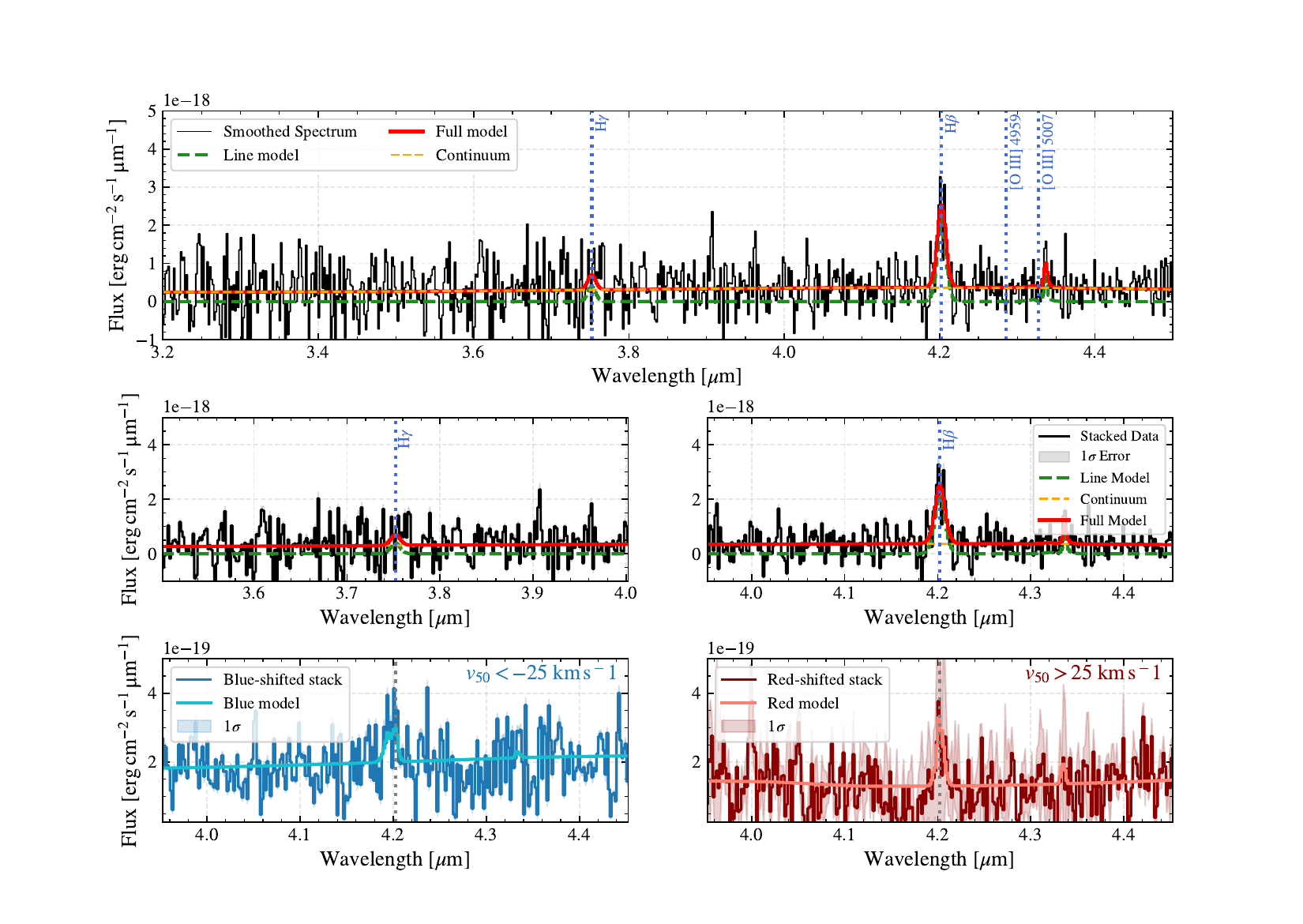}
    \caption{\textit{Upper panel}: Mean spectrum of the Balmer shell. This spectrum was extracted from 57 spaxels with at least $3\sigma$ \hbetashort detections. The line (green dashed line), the continuum (orange dashed line), and the full (red line) models are from our \texttt{q3dfit} analysis. The fitted, redshifted \oiii and continuum emissions are consistent with noise.
    \textit{Middle panel}: Zoom in spectra of the \hgamma and \hbetashort regions. \textit{Lower panels:} Mean spectra of blueshifted ($v_{50} < -25 \,\mathrm{km\,s^{-1}}$, left) 
and redshifted ($v_{50} > 25 \,\mathrm{km\,s^{-1}}$, right) \hbetashort spaxels.}
    \label{fig:stack}
\end{figure*}

Considering only the 57 spaxels with at least a $3\sigma$ detection of H$\beta$, we constructed a mean stacked spectrum of the shell shown Fig.~\ref{fig:stack}.  The \hbetashort line is clearly detected, while \oiii remains undetected, and with a median S/N$_\mathrm{H\gamma}=1.44$, H$\gamma$ is not detected. We also show the mean stacked spectra, split by their $v_{50}$ values, to highlight differences in the H$\beta$ line profile between blueshifted ($v_{50} < -25 \mathrm{km,s^{-1}}$) and redshifted ($v_{50} > 25 \mathrm{km,s^{-1}}$) regions of the H$\beta$ shell. The blueshifted spaxels exhibit a slightly higher continuum and weaker emission lines compared to the redshifted ones.

We compute the integrated $3\sigma$ upper limit on [O III] $\lambda$5007 in the H$\beta$ shell (57 spaxels with \hbetashort at SNR$>3$) as
\begin{equation}
     F_{\mathrm{[OIII]}<3\sigma}  = 3 \sqrt{\sum \sigma_{ij,\mathrm{[OIII]}}^2},
\end{equation}
where $\sigma_{ij,\mathrm{[OIII]}}$ is the per-spaxel $1\sigma$ noise uncertainty in the spectrum at the location of [O\,\textsc{iii}].

The corresponding integrated flux ratio upper limit is then
\begin{equation}
     \log_{10} \left( \frac{F(\mathrm{[OIII]})}{F(\mathrm{H}\beta)} \right)_{<3\sigma} 
    = \log_{10}\left( \frac{  F_{\mathrm{[OIII]}<3\sigma}}{\sum_{ij} F_{ij}(\mathrm{H}\beta)} \right) 
    = -1.15,
\end{equation}
where $F_{ij}(\mathrm{H}\beta)$ is the spaxel-wise flux measured in the $3\sigma$-detected H$\beta$ line. The integrated flux of \oiii in the extended ionized shell is thus lower than $10^{-1.15}\sim 7.1\%$ the integrated \hbetashort flux at the 3$\sigma$ confidence level. Using the same methodology we also measure the 3$\sigma$ upper limit on the \heii to \hbetashort ratio in the \hbetashort nebula. We obtain  $\log_{10} \left( \frac{F(\mathrm{[HeII]})}{F(\mathrm{H}\beta)} \right) _{<,3\sigma} = -1.15$.

\section{Discussion}

We discuss separately the nuclear (Sec.~\ref{sec:nuclear_discussion}) and kiloparsec-scale (Sec.~\ref{sec:kpc_discussion}) properties derived for J0313$-$1806. We discuss our interpretation of the results as evidence for episodic feedback cycles in Sec.~\ref{sec:cycles}.

\label{sec:discussion}

\subsection{Suppressed nuclear [O{\sc iii}] emission}
\label{sec:nuclear_discussion}

The nuclear quasar spectrum displays a broad \hbetashort\ line and strong Fe~{\sc ii} emission, while \oiiidd\ emission is absent (Fig.~\ref{fig:quasar_spectrum}). Weak or absent \oiiidd\ emission is not uncommon in quasar samples across cosmic time \citep[e.g.,][]{netzer04}. In the following, we discuss similar objects in the literature and possible physical origins for the lack of \oiiidd\ in J0313--1806.

\subsubsection{A high-z weak line quasar?}

Weak-line quasars (WLQs) are AGN with absent or very low equivalent-width broad UV emission lines, most prominently the high-ionization lines (e.g. \civ, \heii). The resonant low-ionization line \lya\ is often also weak, and mid-ionization lines (e.g. \ciii], \siiii]) are frequently damped \citep{fan99,diamondstanic09,shemmer10,plotkin15,chen24}. The number density of WLQs appears to increase with redshift, with a fraction of roughly $10-15\%$ at $z \sim 6$ \citep{banados16,shen19}, compared to $1.3\%$ in the \textit{Sloan Digital Sky Survey} (SDSS; \citealt{york00}) at $z \lesssim 4.2$ \citep{diamondstanic09}.

Broad UV high-ionization line measurements for J0313--1806, in particular \civ, have already been presented by \citet{yang21}.
 They report a \civ\ restframe equivalent width EW $=14.2 \pm 0.9 \,\AA$ and a blueshift of \civ\ relative to the systemic redshift from \cii\ of $\Delta v_{\mathrm{CIV-[CII]}} 
   = -4138 \pm 350 \, \mathrm{km \, s^{-1}}$. This makes the line slightly stronger than the canonical EW $<10 \,\AA$ criterion used to define WLQs by \cite[][see also \citealt{shemmer10,wu11,luo15,plotkin15,chen24}]{diamondstanic09}, but consistent with the EW $<15 \,\AA$ criterion adopted by \citet{ni18,timlin20}. J0313$-$1806 thus 
   lies at the boundary of commonly adopted WLQ defintions and could be considered a weak-line quasar under the more inclusive criterion.

The combined \civ\ equivalent width and strong blueshift of J0313--1806 are consistent with those of WLQs (see Fig.~2 of \citealt{ni18} and Fig.~9 of \citealt{timlin20}), and are well offset from the locus of typical quasars. To quantify this behavior in a way that reduces the degeneracy between \civ\ EW and blueshift, we adopt the CIV $\mid\mid$ \textit{Distance} metric (\citealt{rivera20,rivera22}, computed following \citealt{ha23}). This metric combines the two observables into a single scalar quantity that measures how far a source lies along the empirical \civ sequence defined by the quasar population, with larger values corresponding to more extreme \civ properties. The resulting the CIV $\mid\mid$ \textit{Distance} for J0313–1806 is $\simeq 1.10$. At the independently measured \hbetashort-based accretion rate of $L_{\rm bol}/L_{\rm Edd}=0.80\pm0.05$, this value is consistent with the locus occupied by WLQs in the C IV‖Distance–$L_{\rm bol}/L_{\rm Edd}$ plane (Ha et al. 2023), supporting the classification of J0313–1806 as WLQ-like.

\begin{figure}
    \centering
    \includegraphics[width=0.95\linewidth]{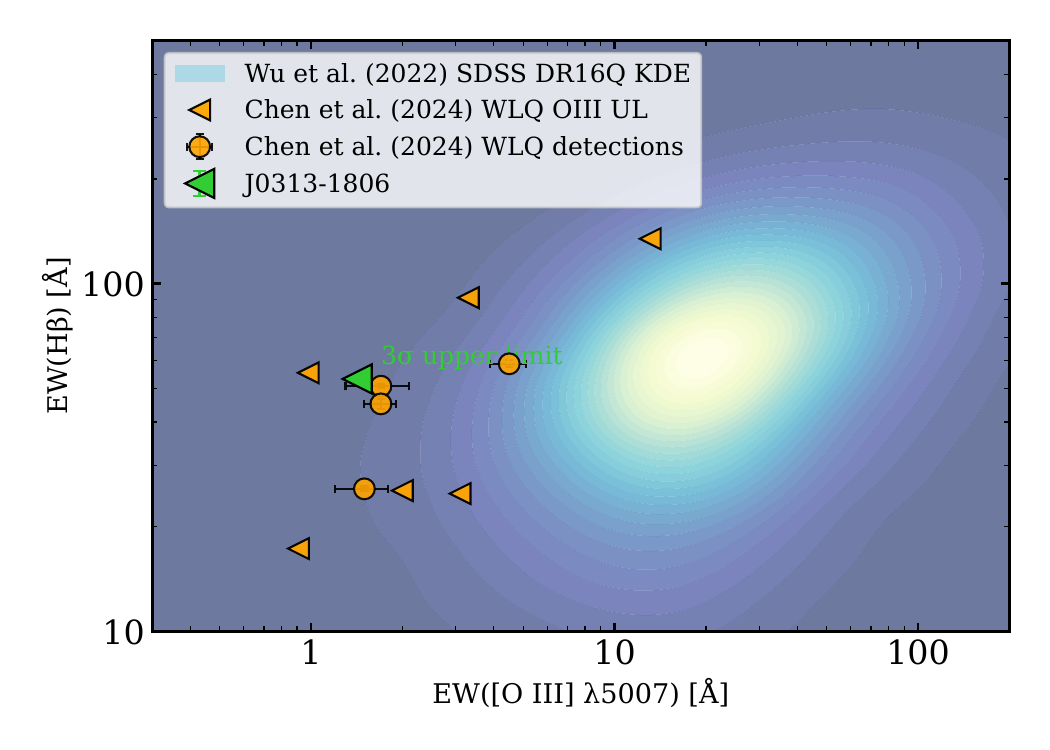}
    \caption{Equivalent widths of \hbetashort\ and upper limit  on \oiii\ for J0313-1806 (green triangle). We compare it to a sample of WLQs identified by \citep{chen24} (orange circles for detected \oiii, orange triangles for upper limits). We also show Gaussian kernel density contours for SDSS DR16Q quasars \citep{wu22}. This highlights the particular weakness of the \oiiidd\ emission in the nuclear spectrum of J0313--1806, even relative to typical WLQs.}
    \label{fig:ew}
\end{figure}

We measured the restframe equivalent widths of \hbetashort\ and the \feii\ in the 4434–-4684\,$\AA$ spectral window from the quasar spectrum extracted in Section~\ref{sec:spec_fit}. The equivalent width of the fitted broad \oiiidd wings is extremely small ($\approx 0.02 \, \AA$) 
and we thus obtained a conservative 3$\sigma$ upper limit for the equivalent width of \oiii\ in the spectral window 4988--5028 $\AA$.  
We obtain EW(H$\beta$)=$53.18\pm 0.31 \, \AA$, EW(\feii)=$38.46\pm 0.02 \, \AA$ and EW(\oiii)< 1.42 $\AA$ (see Table \ref{tab:spec}) and compare these values to the WLQ sample of \citet{chen24}, as well as to a control sample of typical SDSS quasars (135,738 quasars with EW(H$\beta$) and EW([OIII]) values from data release~16, DR16Q, \citealt{lyke20,wu22}) in Fig.~\ref{fig:ew}. 
The \oiiidd\ emission in J0313--1806 is particularly weak relative to its \hbetashort\ emission, even when compared to typical WLQs. 

The physical driver of the UV-line weakness in WLQs remains debated. One possibility is that the BLR is “anemic,” i.e. has a low gas content and/or covering factor \citep{shemmer10}. Alternatively, a soft ionizing continuum could also produce weak high-ionization emission lines. Investigating a sample of seven WLQs, \citet{plotkin15} find that their restframe optical low-ionization lines (e.g. \hbetashort) are not as exceptionally weak as their UV lines. These findings favor the soft-ionizing-continuum scenario, since an anemic BLR would result in all broad lines being similarly weak.

\citet{plotkin15} further report typical \hbetashort\ widths with FWHM $<4000 \,\mathrm{km\,s^{-1}}$, strong Fe~{\sc ii} emission, and large \civ\ blueshifts (up to $5500 \,\mathrm{km\,s^{-1}}$). As many as $\sim 50\%$ of all WLQs are X-ray weak \citep[e.g.,][]{wu11,luo15,ni18,ni22,pu20} relative to typical restframe UV-to-X-ray ratios in Type 1 AGN \citep{steffen06}. The hard X-ray spectra of X-ray weak WLQs (typical photon index range, $\Gamma \sim 1.1$--$1.2$; \citealt{luo15,pu20}) support the idea that absorption and Compton-thick shielding play an important role for the line emission in WLQs (see also \citealt{ni22}).

In the high- or super-Eddington accretion regime, the accretion disk becomes geometrically thick \citep[][see also \citealt{abramowicz05} and references therein]{abramowicz88}. Once puffed up, the disk can shield the BLR from hard ionizing radiation, resulting in an overall soft SED \citep{wang14shadow,madau24,lupi24,madau24}. This shielding can explain both the absence of high-ionization lines in WLQs and their observed X-ray weakness. Additionally, the large \civ\ blueshifts observed in WLQ samples \citep{luo15,plotkin15} are consistent with the presence of powerful line-driven winds expected in high or super-Eddington disks \citep{castor75,murray95}. Another recurring feature is the weakness of \oiiidd\ emission observed in WLQ samples \citep{leighly07,shemmer10,wu11,plotkin15,ha23,eilers23,chen24}. In the puffed-up high-$L/L_\mathrm{Edd}$ scenario, the shielding of $>35 \,\mathrm{eV}$ photons can also naturally explain the absence of \oiiidd\ lines in WLQ spectra. The anti-correlation between the strength of \oiiidd\ emission and the accretion rate has long been studied in the Eigenvector 1 context \citep[EV1, e.g.,][]{boroson92,sulentic00,marziani01,Shen14,wolf20}. In addition, the Baldwin effect \citep{baldwin77,zhang13,stern13,Shen14} describes the general decrease in the equivalent width of emission lines, including \oiiidd, with increasing quasar luminosity. Indeed, weak or absent \oiiidd\ emission has also been observed in targeted luminous quasar samples, such as the WISE/SDSS-selected hyper-luminous (WISSH) quasar survey \citep[][]{vietri18}, which probes the most extreme end of the quasar luminosity function. \citet{vietri18} argue that the relative \oiiidd\ weakness observed in $\sim 70\%$ of their sample is mainly driven by orientation effects. The NLR is photoionized within the polar ionization cone. When viewed more face-on, i.e., looking down the cone, the bright continuum outshines the projected NLR emission, resulting in a higher continuum-to-NLR ratio than in more edge-on orientations. In addition to high $L/L_\mathrm{Edd}$ and orientation effects, obscuration by dust or dense gas has also been invoked to explain reduced \oiiidd\ emission in quasars \citep{temple19,wang25}. 

Another well-known class of fast accretors that display weak \oiiidd\ emission are the Narrow-Line Seyfert 1 galaxies (NLSy1; \citealt{osterbrock85, goodrich89}). However, we note that these objects are strictly classified based on their narrower \hbetashort\ profiles ($<2000 \, \mathrm{km\, s^{-1}}$), a criterion that is not met by J0313--1806.

\subsubsection{J0313--1806 and the quasar main sequence}
\label{sec:ev1}

In addition to sharing characteristic WLQ properties, J0313--1806  is also a BAL quasar, 
indicative of strong outflows and/or shielding on nuclear scales. Such BAL features are extremely rare among WLQs \citep[e.g.,][]{nikolajuk12}, with one other case reported by \citet{yi19}. 
To anchor J0313--1806 in the EV1 context, we place it along the quasar main sequence \citep[see, e.g.,][for a review]{marziani18}, in the primary EV1 plane defined by the FWHM of H$\beta$ and the relative strength of the optical Fe\,{\sc ii} emission, quantified as 
$r_{\mathrm{FeII}} = F(\mathrm{Fe\,II}) / F(\mathrm{H}\beta)$. 
For J0313--1806, we had obtained $\mathrm{FWHM}(\mathrm{H}\beta) = 4156 \pm 68~\mathrm{km\,s^{-1}}$ and compute $r_{\mathrm{FeII}} = 0.95\pm 0.13$. Archetypal quasar categories have been defined along this sequence \citep{sulentic00}, thought to be primarily driven by a combination of accretion rate ($L/L_{\mathrm{Edd}}$) and the viewing angle toward a flattened BLR \citep[e.g.,][]{Shen14}. A canonical boundary at $\mathrm{FWHM}(\mathrm{H}\beta) \simeq 4000~\mathrm{km\,s^{-1}}$ separates rapidly accreting sources (Population~A, $\mathrm{FWHM}(\mathrm{H}\beta) \lesssim 4000~\mathrm{km\,s^{-1}}$) from more massive, lower-accretion systems (Population~B, $\mathrm{FWHM}(\mathrm{H}\beta) \gtrsim 4000~\mathrm{km\,s^{-1}}$) \citep[e.g.,][]{marziani01}. WLQs typically occupy the extreme Population~A (xA) regime, characterized by $r_{\mathrm{FeII}} \gtrsim 1$ \citep{diamondstanic09,martinezadama18,dultzin20}. 

\begin{figure}
    \centering
    \includegraphics[width=\linewidth]{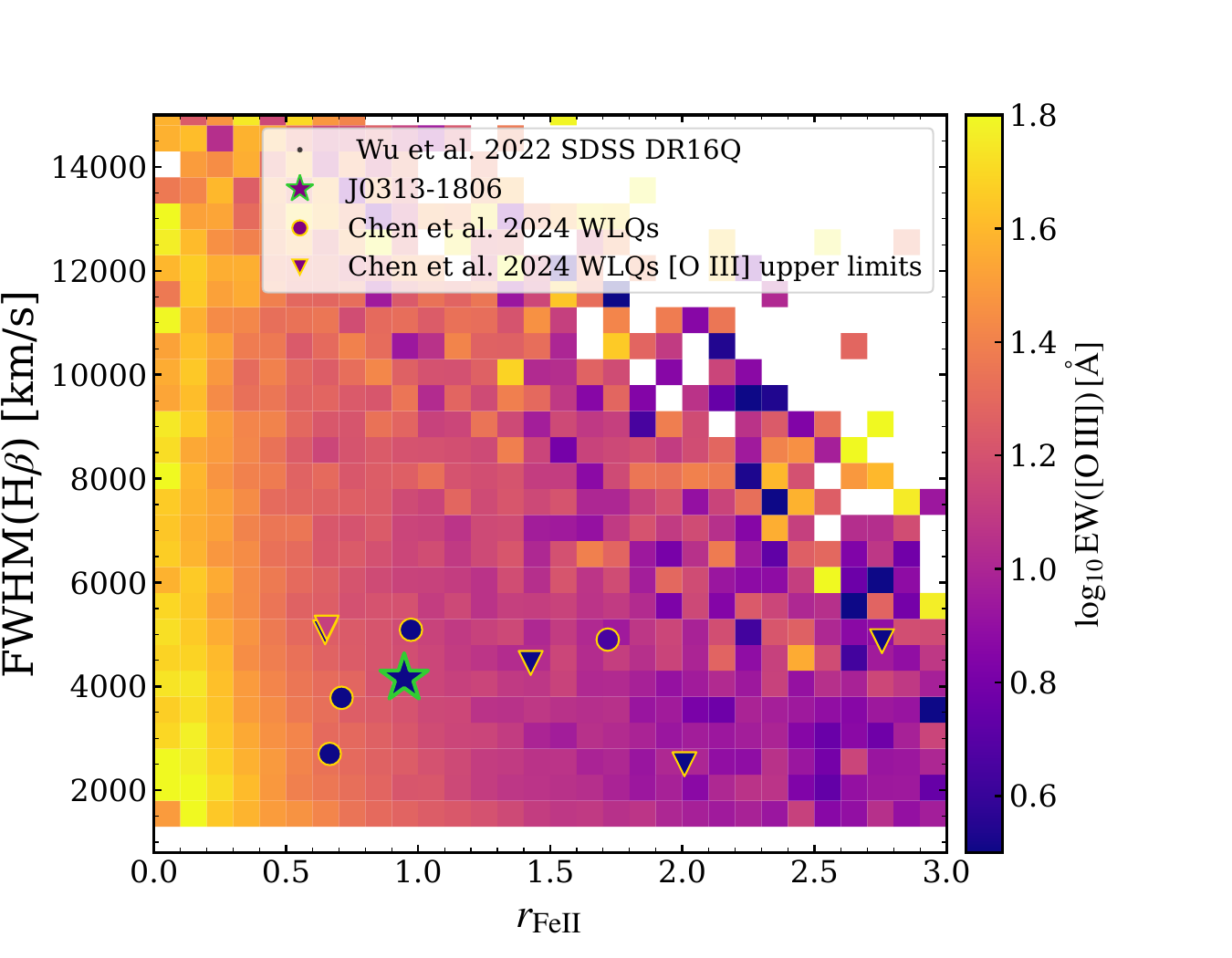}
    \caption{EV1 plane FWHM(H$\beta$) vs. $r_{\mathrm{FeII}}$. The distribution of a subset of SDSS DR16Q quasars from \citet{wu22} is shown in black (see text for selection criteria). The colourmap gives the mean $\log_{10}\,\mathrm{EW}([\mathrm{O\,III}])$ in bins of size $\Delta r_{\mathrm{FeII}} = 0.1$ and $\Delta \mathrm{FWHM}(\mathrm{H}\beta) = 500~\mathrm{km\,s^{-1}}$. J0313-1806 is shown as a green star. WLQs from \citet{chen24} are shown as circles (measured EW(\oiii)) or triangles (upper limits), colour-coded according to their $\log_{10}\,\mathrm{EW}([\mathrm{O\,III}])$. J0313--1806 and several WLQs lie outside the xA regime ($r_{\mathrm{FeII}}>1$) but nonetheless show strikingly weak \oiii\ emission compared to SDSS DR16Q quasars in the same EV1 bins.}
    \label{fig:ev1}
\end{figure}

In Fig.~\ref{fig:ev1}, we show the distribution of SDSS DR16Q quasars \citep{lyke20,wu22} in the redshift range $0.4<z<0.8$, selected following \citet{Shen14} with the criteria $0<r_{\mathrm{FeII}}<3$, \texttt{uni\_flag}$\neq 0$, $800~\mathrm{km\,s^{-1}} < \mathrm{FWHM}(\mathrm{H}\beta) < 15,000~\mathrm{km\,s^{-1}}$, and \texttt{sn\_flag}$>10$. The colour scale indicates the mean $\log_{10}\,\mathrm{EW}$([OIII]) in bins of $\Delta r_{\mathrm{FeII}} = 0.1$ and $\Delta \mathrm{FWHM}(\mathrm{H}\beta) = 500~\mathrm{km\,s^{-1}}$. We mark the position of J0313--1806 and overlay the WLQs presentheted by \citet{chen24}, colour-coded by their measured or upper-limit EW([OIII]). J0313--1806 lies at the edge of the xA regime (formally $r_{\mathrm{FeII}}>1$). Several WLQs from \citet{chen24} 
also occupy a similar region of the plane. Notably, several of these objects exhibit EW(\oiii) values or limits that fall far below the typical SDSS DR16Q quasars in their local EV1 bins. 

Quantitatively, the 3$\sigma$ upper limit on EW(\oiii) for J0313--1806 is a factor of $\sim 37$ lower than the median value of SDSS DR16Q quasars at the same EV1 location (EW([OIII]$)=15.76\,\AA)$, bin $0.90 \le r_{\mathrm{FeII}} < 1.00$ and $3800 \le \mathrm{FWHM}(\mathrm{H}\beta) < 4300~\mathrm{km\,s^{-1}}$, containing 369 sources). This deficit corresponds to a one-sided percentile of $p=3.8 \times 10^{-7}$, i.e. a $4.95\sigma$ outlier with respect to the local quasar population.

 J0313--1806 exhibits exceptionally weak nuclear \oiii\ emission compared to both the general quasar population at the same EV1 location and to local WLQs. This suggests that the extreme weakness of \oiiidd\ in J0313--1806 cannot be attributed solely to high- or super-Eddington accretion, but likely requires additional factors such as orientation, obscuration, or collisional de-excitation. In the paradigm presented by \citet{Shen14}, at a fixed $r_{\mathrm{FeII}}$, the spread in FWHM(H$\beta$) primarily reflects orientation effects, with broader FWHM(H$\beta$) corresponding to more edge-on viewing angles and narrower profiles to more face-on systems. For J0313$-$1806, our measured $\mathrm{FWHM}(\mathrm{H}\beta)$ and $r_{\mathrm{FeII}}$
 place it among the broader \hbetashort profile sources at this $r_{\mathrm{FeII}}$ (see black dots in Fig.~\ref{fig:ev1}), suggesting more edge-on configurations. Consequently, orientation alone is also unlikely to explain the unusually weak \oiiidd emission.

\subsection{Suppressed extended [O{\sc iii}] emission}
\label{sec:kpc_discussion}

To further constrain the scenarios responsible for weak \oiiidd in J0313--1806 we will now discuss the kpc-scaled \hbetashort emitting shell illuminated around the nucleus revealed in our JWST/NIRSpec IFU observation.

\subsubsection{Photoionization simulations}

In AGN, narrow-line emission can be observed out to kpc scales around the central continuum source  \citep[e.g.,][]{bennert02,schmitt03,greene11,hainline14,liu14}. In quasar NLRs typical densities reach $n_{\mathrm{e}} \sim 10^{2-4}\mathrm{cm^{-3}}$ \citep[e.g.,][]{bennert06,nagao06,kakkad18,joh21}, allowing forbidden high-ionization lines such as \oiiidd\ to be efficiently produced in partially ionized gas. 
This is not the picture observed in the extended emission around J0313--1806, where we report broadened \hbetashort emission, with a median velocity dispersion of $\sigma=(415 \pm 51)\, \mathrm{km\, s^{-1}}$ but no significant \oiiidd in the immediate kpc-scale environment of the quasar. 
The observed peak in the \hbetashort\ flux distribution suggests that the ionized shell is illuminated by the central source. The absence of \oiiidd\ emission throughout the shell is intriguing, and we explore potential physical scenarios to explain this in light of the simultaneous lack of nuclear \oiiidd\ emission (Sec.~\ref{sec:nuclear_discussion}).

One of the leading physical explanations for the absence of \oiiidd\ in luminous quasars, as discussed in Section \ref{sec:ev1}, is an orientation effect: the BLR outshines the NLR along low-inclination sightlines \citep[e.g., ][]{vietri18}
. However, the complete \emph{absence} of extended \oiiidd\ emission in the PSF-subtracted NIRSpec/IFU cube disfavors this interpretation for the shell. The extended and nuclear \oiiidd\ weakness must therefore be explained by the physical state of the gas and/or the ionizing SED rather than by orientation alone. 

We carry out simulations with the \texttt{python} front-end \textsc{pyCloudy} \citep[\texttt{v0.9.15};][]{morisset13} and the \textsc{Cloudy} photoionization code \texttt{c23.01} \citep{ferland13,Gunasekera23}, which computes the thermal, ionization, and line-emission structure of the shells. 

\begin{figure}
    \centering
    \includegraphics[width=0.95\linewidth]{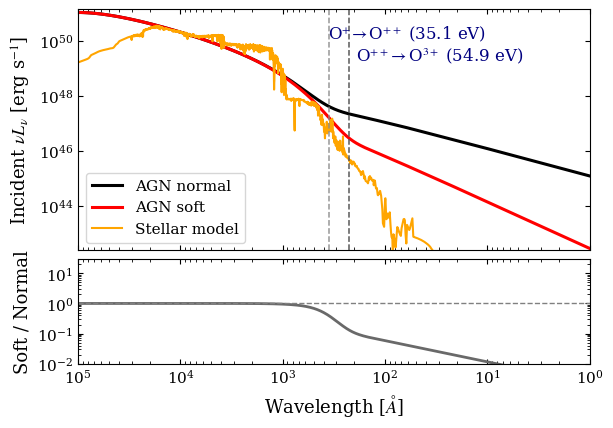}
\caption{AGN and stellar continuum models adopted in the \textsc{Cloudy} simulations. 
\textit{Upper panel:} The black line shows our "AGN normal" model with $T=1.0\times10^{5}\,\mathrm{K}$, $\alpha_{\rm uv}=-1.09$, $\alpha_{x}=-1.0$ ($\Gamma\simeq2.0$), and $\alpha_{\rm ox}=-1.4$, representative of luminous quasars with typical X-ray emission. The red line shows our "AGN soft" model with $T=1.0\times10^{5}\,\mathrm{K}$, $\alpha_{\rm uv}=-1.09$, $\alpha_{x}=-1.7$, and $\alpha_{\rm ox}=-2.2$, chosen to mimic the steeper and X-ray–weaker spectra representative of WLQs. The orange line shows the BPASS starburst model. The vertical dashed lines indicate the ionization thresholds for O$^{++}$ (35.1 eV) and O$^{3+}$ (54.9 eV). 
\textit{Lower panel:} Ratio of the two continua, highlighting that the "AGN soft" model strongly suppresses the hard ionizing photon budget compared to the "AGN normal" case.}

    \label{fig:continua}
\end{figure}

Since \halpha, as well as the forbidden lines \nii\ and \sii, lie outside the spectral window of our observations, and other diagnostic features such as the auroral \oiiiaur \ line are too faint to be detected, we cannot apply standard emission line ratio diagnostics \citep[e.g][]{baldwin81,veilleux87,kauffmann03,kewley06,mazzolari24,scholtz25}. We use two primary diagnostic ratios :$\log_{10}([\mathrm{O\,III}]\,5007/\mathrm{H}\beta)$ and $\log_{10}([\mathrm{He\,II}]\,4686/\mathrm{H}\beta)$.
We model a spherical shell of photoionized gas extending from $r_{\rm in}\in\{0.1,\,0.8,\,1.7\}$\,kpc to $r_{\rm out}=1.8$\,kpc, with covering factor $C_f=0.9$. The covering factor mimics an opening along our line of sight, exposing the BLR, as required by the observed nuclear broad H$\beta$.  We note that varying $C_f$ rescales the total line luminosities but leaves ratios unchanged to first order. The gas is dust-free, has constant density at the illuminated face, and is spherical (\texttt{sphere}). We vary metallicity over $\log(Z/Z_\odot)\in[-2.50,\,0.25]$ in steps of 0.25 dex and density a wide range $\log n_{\rm H}/\mathrm{cm}^{-3}=\{3-8 \}$ in steps of 0.5 dex. Metallicity-dependent elemental abundances are adopted following the analytic prescriptions of \citet[][see also \citealt{decarli24}]{nicholls17}; in particular the oxygen abundance scales as
\begin{equation}
    12 + \log(\mathrm{O}/\mathrm{H}) \;=\; 8.76 + \log Z,
\end{equation}
with $\log Z$ the logarithmic metallicity relative to solar. We bracket the possible ionizing spectral shapes with two parameterized AGN continua using \textsc{Cloudy}'s \texttt{agn} command.
The rest–UV spectral slope is fixed by the spectrum in  \cite{wang21}: 
$\alpha_\lambda \simeq -0.91$ near Ly$\alpha$, which implies 
$\alpha_\nu = -\,(2+\alpha_\lambda) = -1.09$. 
We therefore hold $\alpha_{\mathrm{uv}}=-1.09$ in all models and vary only the high-energy shape of the SEDs. 
Our "AGN normal" continuum adopts $T = 1.0\times10^{5}\,\mathrm{K}$, $\alpha_{\mathrm{uv}}=-1.09$, $\alpha_{x}=-1.0$ ($\Gamma\simeq2.0$), and $\alpha_{\mathrm{ox}}=-1.4$, 
representative of luminous quasars with standard X-ray output relative to their optical emission 
\citep{steffen06,just07,lusso10,nanni17,vito19}. 
The "AGN soft" (WLQ-like) continuum instead has $T = 1.0\times10^{5}\,\mathrm{K}$, $\alpha_{\mathrm{uv}}=-1.09$, $\alpha_{x}=-1.7$, and $\alpha_{\mathrm{ox}}=-2.2$, 
mimicking an extremely X-ray–weak quasar similar to observed WLQs 
\citep[e.g.,][]{luo15,timlin20}. 
Both continua are normalized to the observed optical luminosity 
$\nu L_\nu(5100\,\text{\AA}) = 1.77\times10^{46}\,\mathrm{erg\,s^{-1}}$, as derived in Section~\ref{sec:spec_fit}. 
 The source models are shown in Fig. \ref{fig:continua}. Line emissivities are extracted per zone and integrated to total line powers.
The O$^{+}\rightarrow$O$^{++}$ ionization edge lies at $35.1$ eV ($354$ \AA); the suppression of photons above this threshold, like in our soft model naturally reduces the production of O$^{++}$ and thus the strength of \oiiidd emission. However, in harder ionizing spectra, photons beyond the ionization edge at 54.9 eV (O$^{++}\rightarrow$O$^{3+}$) further ionize oxygen, generating $\mathrm{O}^{3+}$ and depleting \oiiidd. We aim to test the net effect of the EUV-X-ray SEDs on the O$^{++}$ ion population with our two models.

We further set up a "nuclear star formation" ionizing source model using a \textsc{BPASS} v2.2.1 \citep{eldridge17,stanway18} spectrum corresponding to a 10~Myr instantaneous burst at solar metallicity (orange line in  Fig.~\ref{fig:continua}). The spectrum was normalized to a total ionizing photon output of 
$\mathrm{log_{10}} \, Q(\mathrm{H}) \sim 50$, 
consistent with a vigorous nuclear starburst producing several~$\times\,10^{2}\,M_{\odot}\,\mathrm{yr^{-1}}$ \citep[e.g.,][]{kennicutt98}
typical of luminous $z\sim7$ quasar hosts. We stress, however, that the quasar should overwhelmingly dominate the ionizing photon budget. Nuclear star-formation is included only as a comparison case to illustrate the ionization conditions expected in the absence of a hard AGN spectrum, rather than as a realistic alternative single power source.

\begin{figure*}
    \centering
    \includegraphics[width=0.80\linewidth]{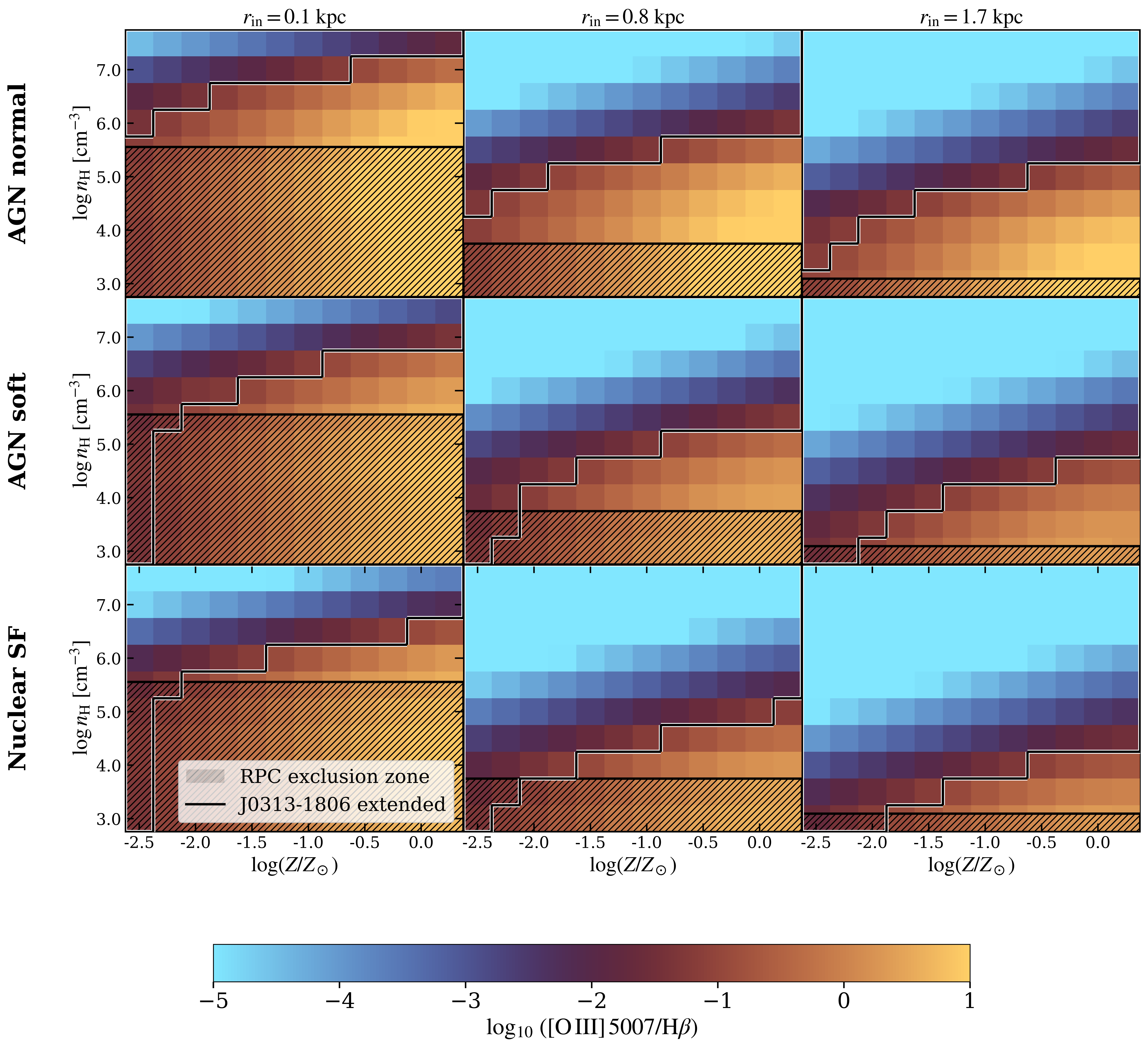}
    \caption{Cloudy simulation results for the H$\beta$ shell. The upper and middle rows correspond to models using the AGN normal" and AGN soft" continua as ionizing sources, respectively, while the lower row corresponds to the stellar ionizing source. Each panel shows the predicted line ratio  $\log_{10}([\mathrm{O\,III}]\,5007/\mathrm{H}\beta)$ as a function of gas density $n_{\rm H}$ and metallicity $Z$, for inner shell radii $r_{\rm in}=0.1$, $0.8$, and $1.7$ kpc (left to right). The maps are spanned by constant-density shell models with outer radius $r_{\rm out}=1.8$ kpc. Regions enclosed by the black contour are consistent with the observed constraint $\log_{10}([\mathrm{O,III}]/\mathrm{H}\beta)<-1.15$ from the extended emission in J0313-1806. The hatched areas indicate the radiation–pressure confinement (RPC) exclusion zone, where densities fall below the expected RPC floor and are therefore disfavored.}
    \label{fig:models}
\end{figure*}

In Fig.~\ref{fig:models} we show the simulated line ratios  $\log_{10}([\mathrm{O\,III}]\,5007/\mathrm{H}\beta)$ across grids of hydrogen density $n_{\rm H}$ and metallicity $Z$ for the AGN normal", "AGN soft" and stellar models, each evaluated at three inner shell radii $r_{\rm in}\in{0.1,0.8,1.7} \,\mathrm{kpc}$.
Regions where $\log_{10}([\mathrm{O\,III}]5007/\mathrm{H}\beta)<-1.15$ (Section \ref{sec:moment_maps}) are highlighted as consistent with the ratio measured in the integrated spectrum of the shell.
The illuminated skin of an AGN-irradiated shell is expected to be compressed until the gas pressure balances the incident radiation pressure—so-called radiation-pressure confinement (RPC; e.g., \citealt{draine11,stern14III,stern14I,baskin14II,baskin14IV,baskin21}).
This sets a characteristic density floor,
$n_{\rm floor}\sim P_{\rm rad}/(kT) \approx L/(4\pi r^{2} c\,kT)$.
We do not enforce an RPC in our \textsc{Cloudy} models; densities below this floor are shown for completeness but are treated as physically disfavoured.
Figure~\ref{fig:models} shows the corresponding “RPC exclusion zone.” Cosmological zoom-in simulations \citep{costa15} indicate that cold gas can further be pressure-confined by a hot, volume-filling shocked medium \citep[see also][]{stern16}. Even if this hot phase remains undetected in our IFU data, it would only add to the total pressure, so the RPC floor remains a valid lower bound.

We see only minor differences between the "AGN normal" and "AGN soft" continuum-driven runs in terms of the accessible parameter space, with slightly lower densities accessible in the "AGN soft" model. This indicates that, assuming AGN photoionization, the depletion of \oiiidd is primarily driven by collisional de-excitation or metallicity rather than by spectral hardness.  
For all three inner radii probed, there exist regions of parameter space that are consistent with the observational constraint $\log_{10}([\mathrm{O\,III}]\,5007/\mathrm{H}\beta)<-1.15$, as well as regions that are inconsistent. To first order, our models show that reproducing the observed weakness of \oiiidd requires either very high gas densities ($n_{\rm H}\gtrsim10^{6}\,\mathrm{cm^{-3}}$, where collisional suppression becomes efficient independent of metallicity) or, at lower densities ($n_{\rm H}\lesssim10^{5}\,\mathrm{cm^{-3}}$), sub-solar metallicities ($Z\lesssim0.2\,Z_\odot$). Thus, both high-density and low-metallicity solutions are supported.

At very high densities, $n_{\rm H}\gtrsim 10^{7},\mathrm{cm^{-3}}$, collisional de-excitation is so efficient that the observed ratios can be reproduced almost irrespective of metallicity or shell thickness. At densities $n_{\rm H}\sim 10^{5-7}\mathrm{cm^{-3}}$, the line ratios are increasingly suppressed with decreasing metallicity, $Z\lesssim 0.1$–$0.2,Z_\odot$. For more moderate densities, $n_{\rm H}\lesssim 10^{4.5},\mathrm{cm^{-3}}$, the predicted ratios exceed the observed limit across essentially the entire metallicity range. A thinner shell, i.e. with an larger inner radius (e.g. $r_\mathrm{in}=$ 1.7 kpc) significantly relaxes the density requirements, enabling $n_{\rm H}\sim 10^{5}\mathrm{cm^{-3}}$.

The even softer "nuclear star formation" SED enables access to consistent $\log_{10}([\mathrm{O\,III}]\,5007/\mathrm{H}\beta)$ ratios over a larger set of density and metallicity combinations, with $n_{\rm H}\sim 10^{4}\mathrm{cm^{-3}}$ densities available up to log($Z/Z_\odot \sim 1$).

We note, however, that even for the lower possible densities $n_{\rm H}\gtrsim 10^{3.5}\,\mathrm{cm^{-3}}$, the total gas mass implied by a uniform, kpc-scale shell becomes unrealistically large. The ionized gas mass of a shell with inner radius $r_{\rm in}$, outer radius $r_{\rm out}$, covering factor $C_{\rm f}$, and filling factor $f$ is:

\begin{equation}
    M_{\rm gas} \;\approx\; 4\pi\, C_{\rm f}\, f\, \mu m_{\rm p}\, n_{\rm H}\, (r_{\rm out}^{3}-r_{\rm in}^{3})/3,
\end{equation}
where $\mu\simeq 1.4$ is the mean particle mass per hydrogen nucleus. 

Considering the optimistic case of a thin shell with $r_{\rm in}\sim 1.7$ kpc, a shell thickness of 100 pc, $C_{\rm f}=0.9$ and $n_{\rm H}\sim 10^{3.5}\,\mathrm{cm^{-3}}$, this expression yields $M_{\rm gas}\sim 1.3 \times  10^{11}\,M_\odot$ if the gas is volume-filling, i.e., $f=1$. Such a mass is implausibly large for kpc-scale gas reservoirs in early galaxies. Compared to observational constraints from dust and \cii ($M_\mathrm{dust}\sim 7 \times 10^7 \, M_\odot$, \citealt{wang21}) this would require unreasonably high gas-to-dust ratios \citep[realistic gas-to-dust ratios are of the  $10^2$, e.g., ][]{li19}. The dense phase required to collisionally quench \oiiidd cannot be distributed in a volume-filling layer. Instead, the emitting material must occupy only a small fraction of the swept-up volume, implying a low filling factor ($f \lesssim 0.05$) to keep the total ionized mass within reasonable bounds. One plausible realization is emission from dense clumps or a thin layer within the shell, but other geometries with similarly low filling factors cannot be excluded. We are assuming a spherical symmetric shell for simplicity, while the observed distribution is also consistent with a partial shell (or even more complex geometry), which would allow us to reduce the total gas mass by a factor of up to $\sim 10$.

Our observed upper limit in [OIII]/\hbetashort ratios is thus consistent with ionization from an AGN or nuclear star formation, only if we assume a clumpy geometry for the shell. In Section~\ref{fig:moment_maps}, we also derived an upper limit on a second line ratio: $\left[ \log_{10} \left( \frac{F(\mathrm{[HeII]})}{F(\mathrm{H}\beta)} \right) \right]_{<,3\sigma} = -1.12$. This ratio is primarily sensitive to the hardness of the ionizing radiation field and can be used to distinguish between AGN- and star formation–driven photoionization scenarios \citep[e.g.][]{decarli24}. However, when comparing our $3\sigma$ upper limit to the simulation results presented by \citet[][see their Fig.~9]{decarli24}, we find that our measurement does not allow us to conclusively differentiate between the two photoionization scenarios. We, however, note that a soft ionizing continuum such as the one from nuclear star-formation model would significantly relax the density, thickness, and filling-factor requirements.

\subsubsection{H$\beta$ shell energetics}

\begin{figure}
    \centering
    \includegraphics[width=0.95\linewidth]{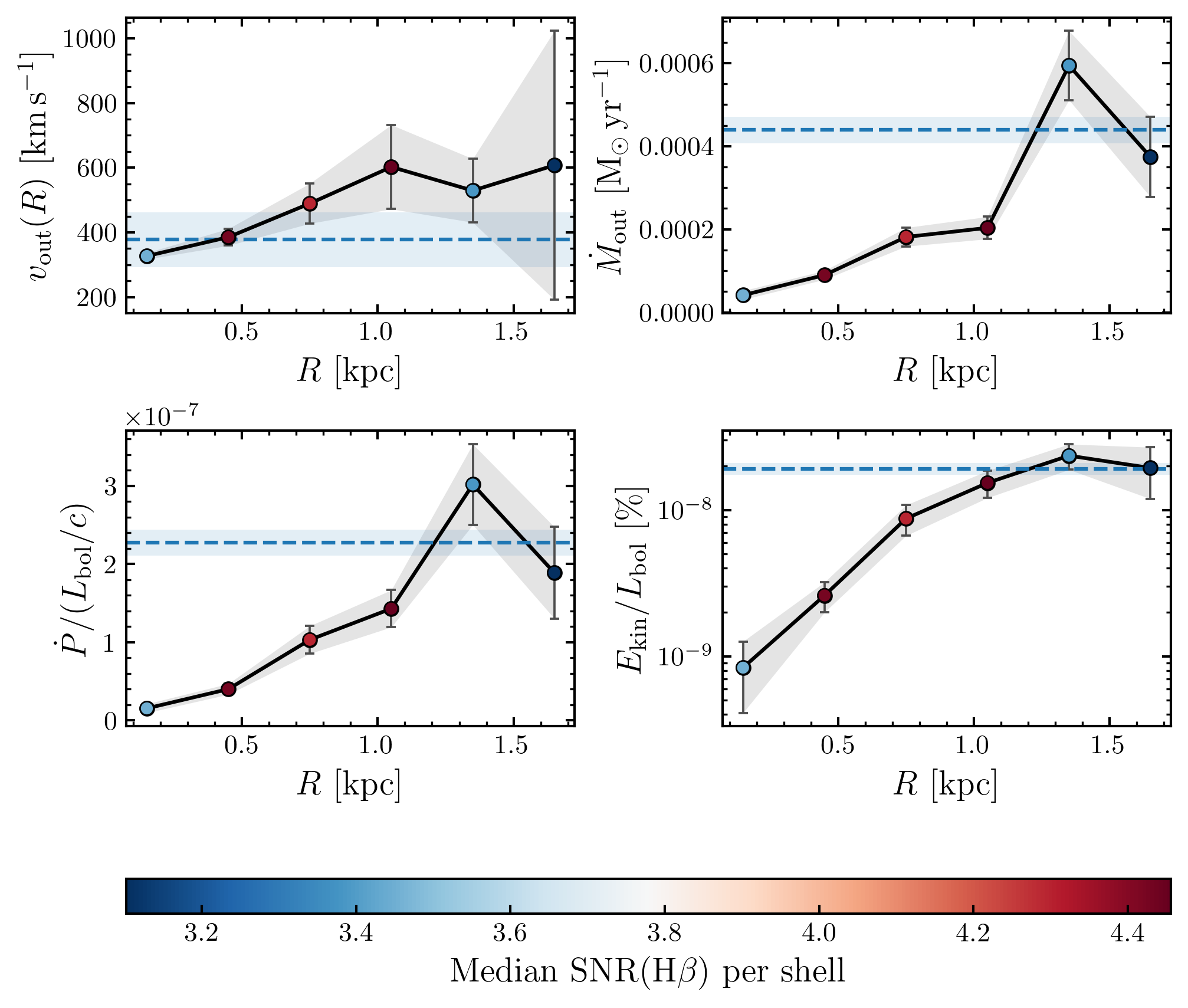}
    \caption{\hbetashort spatially resolved energetics. Here we show the median values of the instantaneous energetics calculated in concentric shells centred on the quasar (circles) and their associated $1\sigma$ uncertainties. The points are colour-coded by median S/N$_\mathrm{H\beta}$ in that shell. The upper right panel shows the outflow velocity, the upper right one shows the mass outflow rate, the lower right one the momentum flux ratio and the lower right one the energy flux ratio. The horizontal lines and shaded areas represent the corresponding integrated measurements and their $1\sigma$ uncertainties. For $v_\mathrm{out}$ we show the median velocity and the median absolute deviation.}
    \label{fig:shells_energ}
\end{figure}

Following \citet{osterbrock06} (see also \citealt{vayner24,liu24}), the recombination mass of the spatially resolved \hbetashort shell contained in each spaxel can be calculated as: 

\begin{equation}
    M_{ion}^{ij} = \frac{m_p \; L_\mathrm{H\beta}^{ij}}{\alpha_\mathrm{H\beta}\;  h\nu_\mathrm{H\beta} \, n_e},
\label{eq:gasmass}
\end{equation}

where $m_p$ is the mass of the proton, $L_\mathrm{H\beta}$ the luminosity of \hbetashort, $\alpha_\mathrm{H\beta}$ is the effective \hbetashort recombination coefficient, $h$ is the Planck constant, $\nu_\mathrm{H\beta}$ is the restframe frequency of \hbetashort and $n_e$ is the electron density. Here we assume that \hbetashort is mainly produced through recombination and model the emitting gas as an ensemble of identical, unresolved gas elements (hereafter referred to as “clouds”). Each cloud has a fixed electron density. Assuming Case B recombination, we calculate the recombination coefficient of \hbetashort using the nebular emission line analysis package \texttt{pyneb} \citep{luridiana15}. 
We note that the recombination mass is formally different than the total gas mass calculated in Eq. \ref{eq:gasmass}, as it only traces the H{\sc ii} region actively emitting \hbetashort.

We consider conditions similar to the ones supported by our \texttt{Cloudy} AGN simulations and set   $n_e=10^6\; \mathrm{cm^{-3}}$
($\alpha_\mathrm{H\beta}=3.071\times 10^{-14} \, \mathrm{cm^{-3} \, s^{-1}}$) and fix the temperature to $T=10^4$ K. These numbers yield a total recombination mass of $M_{ion,\mathrm{tot}}=(1.02 \pm 0.09)\times 10^3 \, M_\odot$ for the shell. 

We can then compute integrated and instantaneous energetics of the outflow. We start by the integrated ionized mass outflow rate:

\begin{equation}
     \dot{M}_{ion}^{int} =  \sum_{ij} \frac{M_{ion}^{ij}\; v_\mathrm{out}^{ij}}{R^{ij}}, 
\end{equation}

where $v_\mathrm{out}^{ij}= \mid v_\mathrm{50}^{ij}\mid  + \sigma^{ij}$ is the outflow velocity, i.e. the sum of the absolute velocity offset traced by the non-parametric $v_\mathrm{50}$ and the velocity dispersion. This convention allows us to capture the maximum outflow velocity encoded in the winds of the emission line \citep{rupke13,vayner24}. $R^{ij}$ represents the physical projected distance between the spaxel $(i,j)$ and the quasar coordinates (49,46). Following \citet{vayner24} we also obtain the spaxel-wise instantaneous mass outflow rate from:

\begin{equation}
     \dot{M}_{ion}^{ij} =  \frac{M_{ion}^{ij}\; v_\mathrm{out}^{ij}}{\mathrm{d}R}, 
\end{equation}
where $\mathrm{d}R$ is the physical size of a single spaxel. We further calculate integrated and instantaneous momentum and kinetic luminosities using the following equations:

\begin{equation}
\begin{aligned}
    \dot{P}_{\mathrm{ion}}^{ij} &= M_{\mathrm{ion}}^{ij} \; v^{ij} \qquad
    E_{\mathrm{kin,ion}}^{ij} = \frac{1}{2} M_{\mathrm{ion}}^{ij} \; (v^{ij})^2
\end{aligned}
\end{equation}

In the absence of a significant detection of H$\gamma$ in the PSF-subtracted cube, we can only place an upper limit on the flux ratio $F(\mathrm{H}\gamma)/F(\mathrm{H}\beta)$ by considering spaxels with robust H$\beta$ detections ($\mathrm{S/N}_{\mathrm{H}\beta}>3$). The resulting limits yield a median constraint of $F(\mathrm{H}\gamma)/F(\mathrm{H}\beta) < 0.44$. This limit is only slightly lower than the theoretical Case~B recombination value of $F(\mathrm{H}\gamma)/F(\mathrm{H}\beta)=0.47$ for $T{\mathrm{e}}=10^{4}$~K and $n_{\mathrm{e}}=100~\mathrm{cm^{-3}}$, and therefore does not provide a meaningful constraint on the reddening. We therefore do not apply an extinction correction to the H$\beta$ flux, noting that any unaccounted extinction would increase the inferred line luminosity and associated energetics. 

 Following \cite{vayner24}, we also measure the median instantaneous energetics in concentric shells around the quasar position with a radius increment of $0.3 \, \mathrm{kpc}$. Our integrated and cumulative measurements are shown in Fig.~\ref{fig:shells_energ}. 
We find that the median instantaneous outflow velocity $v_\mathrm{out}$ increases with radius, reaching $600$--$700 \,\mathrm{km \, s^{-1}}$ at $\sim 1.1 \,\mathrm{kpc}$ and remaining high at larger radii. The overall median value of $377 \pm 85 \,\mathrm{km \, s^{-1}}$ is lower than the median instantaneous velocities measured beyond $1 \,\mathrm{kpc}$, showing that a subset of outer pixels drives the fastest outflows and dominates the kinematics. The instantaneous mass outflow rate $\dot{M}_\mathrm{out}$ and the momentum flux $\dot{P}$ also rise toward larger radii, and their localized peak is consistent with a thin-shell geometry.  

Assuming a high gas density of $n_\mathrm{H}=10^6 \,\mathrm{cm^{-3}}$, the outflow rates remain modest, with an integrated value of $\dot{M}^{\mathrm{int}}_{\mathrm{ion}}=(4.3 \pm 0.3)\times 10^{-4} \,M_\odot \,\mathrm{yr^{-1}}$. The corresponding integrated momentum flux is low, $\dot{P}^{\mathrm{int}}_{\mathrm{ion}}/(L_\mathrm{bol}/c)=(2.3 \pm 0.2)\times 10^{-7} \ll 1$, showing that the outflow is strongly momentum starved compared to the quasar luminosity. The kinetic power of the outflow, both instantaneous and integrated, increases smoothly with radius but remains far below the percent-level energy coupling typically associated with quasar feedback, demonstrating that it is energetically negligible.  

The narrow radial localization of the maximum velocity and mass outflow rate supports the geometry of a thin expanding shell. The very low efficiency relative to $L_\mathrm{bol}$ suggests that the observed ionized component may represent only a thin skin on the surface of a much larger neutral or molecular outflow that is not traced with the NIRSpec/IFU data. The low coupling efficiency might suggest that the driving phase of this outflow has ended, i.e., that it is a fossil remnant of dense gas pushed out earlier in the quasar lifecycle. Assuming a maximum outflow velocity of $\sim 600 \, \mathrm{km\,s^{-1}}$ and a radius of roughly 2 kpc, we obtain $t_{dyn}\sim 3$ Myr, setting a rough timescale for a potential blowout event. High-resolution imaging of the cold molecular phase could reveal whether the ionized outflow is indeed the membrane of a larger molecular region carrying more mass and momentum.

\subsection{Obscured black hole growth and fossil evidence for blowout phase}
\label{sec:cycles}

The recent JWST discovery of high-redshift AGN exhibiting extreme Balmer breaks \citep[e.g.,][]{degraaff25, naidu25} has prompted the development of models of black hole growth deeply embedded in extremely dense gas envelopes \citep[e.g.,][]{inayoshi25, rusakov25, kido25}. These models naturally account for the observed SED shapes without invoking stellar populations.

These findings support the early super-Eddington growth of black holes embedded in dense gas reservoirs \citep[e.g.][]{volonteri10c}. If such an enshrouded phase of black hole growth exists, it must be incorporated into the overall AGN evolutionary cycle and linked to the subsequent luminous quasar phase. Simulations by \citet{lupi24a,husko25} support this type of evolutionary phase. Feedback processes are expected to clear out or consume the surrounding gas during a brief blowout stage \citep[e.g.][]{zakamska16,vayner25}. In cosmological zoom-in simulations presented by \citet{quadri25} the strong feedback event punctuating the rapid accretion phase of SMBHs are followed by a transient quenched phase.

Quasars are indeed expected to undergo most of their early growth phases enshrouded in dust and gas \citep{jahnke20}, with $>82 \%$ predicted to be obscured at $z>7$ \citep[e.g.,][]{davies19}. The recent discovery of blazars at $z\sim 7$ provides further support for prolonged obscured growth phases in early SMBHs \citep{belladitta20,wolf24,ighina24,banados24}.    

Dust obscuration by a nuclear parsec-scale distribution is a key ingredient in the AGN unification model \citep[][for reviews]{urry95,netzer15}. However, the increase of cold gas content in the interstellar medium (ISM) with redshift \citep[e.g.,][]{scolville17,tacconi18} and the simultaneous overall decrease of galaxy sizes \citep[e.g.,][]{allen17,miller07} toward earlier cosmic time has motivated models in which the dense ISM itself can act as an effective obscuring screen \citep{gilli22}. Indeed, combining ALMA and \textit{Chandra} observations of Compton-thick quasars, \citet{gilli14,circosta19,damato20} linked the observed X-ray absorption to the high ISM column densities in their host galaxies.  

Our analysis provides evidence for a thin, clumpy, energetically decoupled ionized shell surrounding J0313$-$1806. One possible interpretation is that this shell is the fossil signature of a recent clearing feedback event (i.e., a blowout), consistent with the strong nuclear outflows reported in the unresolved central region \citep{wang21}. The WLQ-like nuclear properties of J0313$-$1806, in particular the observed \oiiidd\ depletion, are therefore unlikely to be driven primarily by the ionizing continuum shape or orientation. Instead, we propose that the density of the irradiated (and recently expelled) gas slab drives the collisional de-excitation of \oiiidd.

Interestingly, the only two other quasars currently known at $z \sim 7.5$, J100758.26+211529.2 \citep{yang20b,liu24} and ULAS J134208.10+092838.61 \citep[][]{banados18,trefoloni25}, also exhibit relatively weak \oiiidd\ emission and similar \hbetashort\ line profiles in their restframe optical nuclear spectra observed with JWST/NIRSpec IFU. This raises the question of whether similar physical processes operate on their nuclear scales as in J0313--1806, and whether they might likewise host dense, clumpy ionized gas distributions on kiloparsec scales that link WLQ-like behaviour to the obscured phases of quasar evolution. However, the situation may be more complex, as both of these quasars also show kpc-scale \oiiidd\ emission \citep{liu24,trefoloni25}, unlike J0313--1806.

\section{Conclusions}
\label{sec:conclusions}
Our JWST/NIRSpec IFU observations of J0313--1806, the most distant currently known quasar at $z=7.6423$, reveal intriguing nuclear and extended (kpc scale) properties.   
Our findings show that J0313--1806 is a rare BAL WLQ with extraordinarily weak \oiiidd emission. More remarkably, we discovered an extended ionized gas shell spanning kpc scales that completely lacks \oiiidd\ emission yet shows clear \hbetashort. 
The ordered kinematics of this shell are energetically decoupled from the quasar's bolometric output and are likely not driven by ongoing feedback. Our data are consistent with a thin, clumpy-shell geometry, which we interpret as a fossil remnant of a recent blowout phase in the context of obscured black hole growth. If correct, the uncovered shell could represent the ionized skin of a much more massive, cold gas layer.

This interpretation suggests that typical properties of WLQs across cosmic time—\oiiidd\ and X-ray weakness—can be explained by extremely dense interstellar gas with high column densities, which suppresses high-ionization emission lines via collisional de-excitation while absorbing the X-ray emission. In this picture, WLQs represent a transitional stage in which recently cleared, dense, clumpy ISM gas still regulates the observed quasar spectral properties. 

Recent support for this scenario comes from \citet{durovcikova25}, who presented NIRSpec IFU spectroscopy of the $z = 5.9$ WLQ SDSS~J1335+3533. Similar to J0313$-$1806, its nuclear spectrum exhibits weak \oiiidd\ emission potentially blended with the \feii\ pseudo-continuum,  and shows a pronounced weakness of extended \oiiidd\ emission relative to Balmer lines. These striking similarities suggest that the phenomena we observe in J0313$-$1806 may represent a broader evolutionary phase in quasar-host co-evolution.

J0313$-$1806 thus provides an unprecedented view into SMBH-host coevolution at cosmic dawn, establishing a framework for understanding WLQ properties that can now be tested with systematic JWST IFU surveys of early quasars.

\begin{acknowledgements}

 J.W. gratefully acknowledges Hannah Übler for insightful discussions on integral-field spectroscopy at high redshift, and Chris Done for valuable guidance on weak-line quasars and line-driven outflows.
\newline
C.M. acknowledges support from Fondecyt Iniciacion grant 11240336 and the ANID BASAL project FB210003.
\newline

D.S.N.R. was supported under program \#DD-ERS-1335 through a grant from the Space Telescope Science Institute, which is operated by the Association of Universities for Research in Astronomy, Inc., under NASA contract NAS 5-03127.
\newline

J.-T.S. is supported by the Deutsche Forschungsgemeinschaft (DFG, German Research Foundation) - Project number 518006966.

\newline

 B.T. acknowledges support from the European Research Council (ERC) under the European Union's Horizon 2020 research and innovation program (grant agreement number 950533), and by the Excellence Cluster ORIGINS which is funded by the Deutsche Forschungsgemeinschaft (DFG, German Research Foundation) under Germany's Excellence Strategy - EXC 2094 - 390783311.
\newline

RD acknowledges support from the INAF GO 2022 grant ``The birth of the giants: JWST sheds light on the build-up of quasars at cosmic dawn'', the INAF mini-grant 2024: ``The interstellar medium at high redshift'', and the PRIN MUR ``2022935STW'', RFF M4.C2.1.1, CUP J53D23001570006 and C53D23000950006.

\newline

This work is based on observations made with the NASA/ESA/CSA James Webb Space Telescope. The data were obtained from the Mikulski Archive for Space Telescopes at the Space Telescope Science Institute, which is operated by the Association of Universities for Research in Astronomy, Inc., under NASA contract NAS 5-03127 for JWST. These observations are associated with program \#1764.

\newline

The data underlying this article are available from the Mikulski Archive
for Space Telescopes (MAST) at \url{https://doi.org/10.17909/0qd3-x220}.
\newline

Funding for the Sloan Digital Sky 
Survey IV has been provided by the 
Alfred P. Sloan Foundation, the U.S. 
Department of Energy Office of 
Science, and the Participating 
Institutions. 

SDSS-IV acknowledges support and 
resources from the Center for High 
Performance Computing  at the 
University of Utah. The SDSS 
website is www.sdss4.org.

SDSS-IV is managed by the 
Astrophysical Research Consortium 
for the Participating Institutions 
of the SDSS Collaboration including 
the Brazilian Participation Group, 
the Carnegie Institution for Science, 
Carnegie Mellon University, Center for 
Astrophysics | Harvard \& 
Smithsonian, the Chilean Participation 
Group, the French Participation Group, 
Instituto de Astrof\'isica de 
Canarias, The Johns Hopkins 
University, Kavli Institute for the 
Physics and Mathematics of the 
Universe (IPMU) / University of 
Tokyo, the Korean Participation Group, 
Lawrence Berkeley National Laboratory, 
Leibniz Institut f\"ur Astrophysik 
Potsdam (AIP),  Max-Planck-Institut 
f\"ur Astronomie (MPIA Heidelberg), 
Max-Planck-Institut f\"ur 
Astrophysik (MPA Garching), 
Max-Planck-Institut f\"ur 
Extraterrestrische Physik (MPE), 
National Astronomical Observatories of 
China, New Mexico State University, 
New York University, University of 
Notre Dame, Observat\'ario 
Nacional / MCTI, The Ohio State 
University, Pennsylvania State 
University, Shanghai 
Astronomical Observatory, United 
Kingdom Participation Group, 
Universidad Nacional Aut\'onoma 
de M\'exico, University of Arizona, 
University of Colorado Boulder, 
University of Oxford, University of 
Portsmouth, University of Utah, 
University of Virginia, University 
of Washington, University of 
Wisconsin, Vanderbilt University, 
and Yale University.

    Collaboration Overview
    Affiliate Institutions
    Key People in SDSS
    Collaboration Council
    Committee on Inclusiveness
    Architects
    SDSS-IV Survey Science Teams and Working Groups
    Code of Conduct
    SDSS-IV Publication Policy
    How to Cite SDSS
    External Collaborator Policy
    For SDSS-IV Collaboration Members

\newline
This work made use of Astropy:\footnote{http://www.astropy.org} a community-developed core Python package and an ecosystem of tools and resources for astronomy \citep{astropy13, astropy18, astropy22}. 
\newline

 This work made use of v2.2.1 of the Binary Population and Spectral Synthesis (BPASS) models as described in \citet{eldridge17} and \citet{stanway18}.

\newline

 We thank the referee for a careful review of the manuscript and for constructive comments that improved the clarity and robustness of this work.
\end{acknowledgements}

\bibliographystyle{aa}
\bibliography{bibliography}

\begin{appendix}

\section{Post-processing PSF undersampling with \texttt{WICKED}}
\label{sec:appendix_wicked}

\begin{figure}
    \centering
    \includegraphics[width=0.95\linewidth]{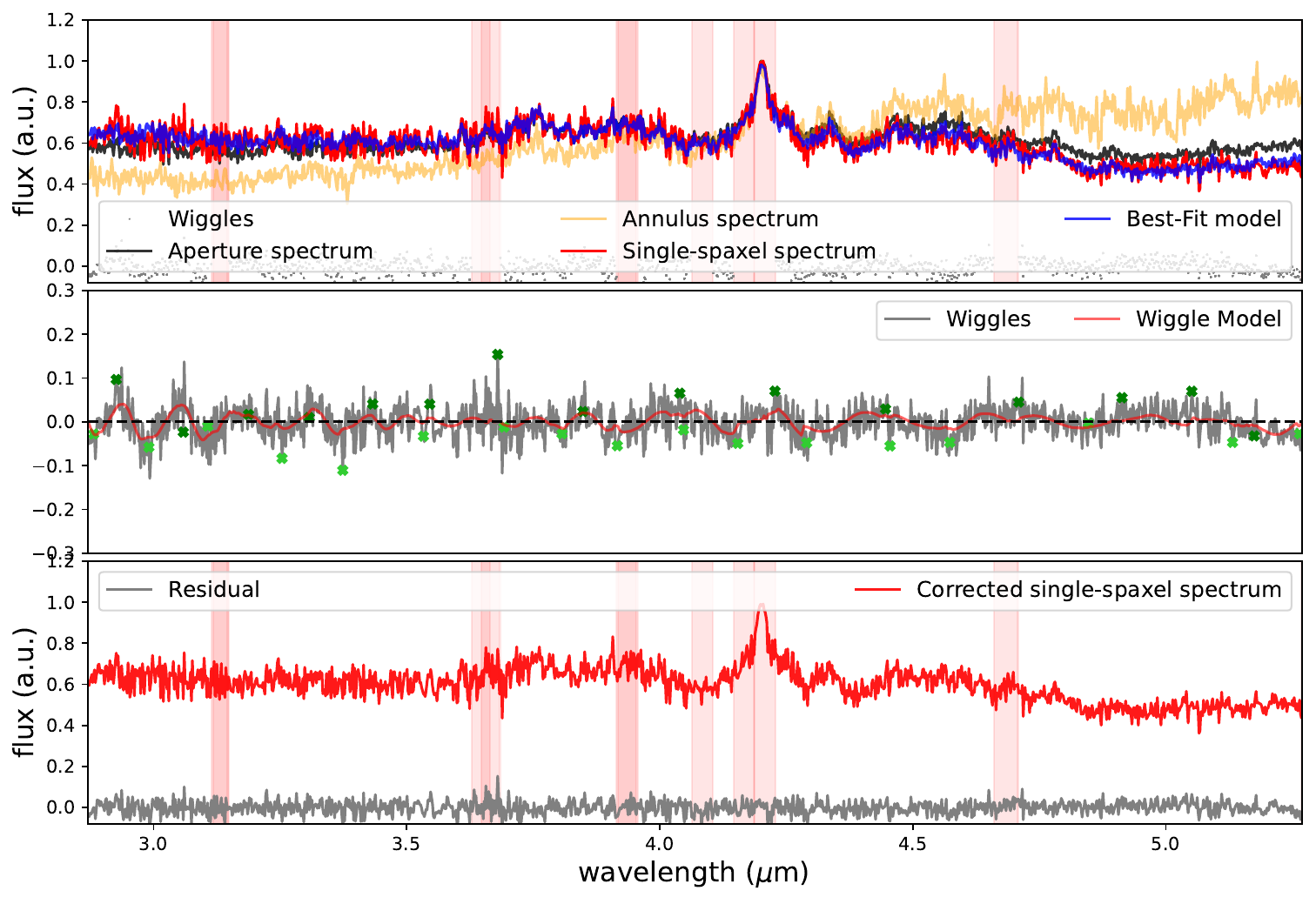}
    \caption{Correction of the brightest pixel of J0313-1806 using \texttt{WICKED}.  \textit{Top:} The solid red line shows the brightest spaxel spectrum of J0313-1806, fitted using a combination of an aperture and annular integrated spectra (black and yellow lines) templates, a power-law and a 2nd-degree polynomial, and shown in blue. Red vertical lines indicate masked regions of known emission lines excluded from the fit. The gray dots display the residuals. \textit{Middle:} The wiggle spectrum is shown in gray and the best-fit wiggle model in red. The green X's are the peaks and valleys of the wiggles spectrum used to divide the wiggle spectrum to constrain its model shown in red (see \citet{dumont25} for details). \textit{Bottom:} In red is shown the wiggle-corrected spectrum, and the residuals in gray.
    }
    \label{fig:wicked}
\end{figure}

Both IFU units of JWST, in particular NIRSpec, are spatially undersampled without reaching Nyquist sampling at any wavelength \citep{law23}. While our dithering strategy partially addresses this issue , we still visually detect sinusoidal modulations or \textit{wiggles} produced by undersampled PSF, as observed in other Quasar data \citep[e.g.][]{perna23,liu24}.
These instrumental artifacts can significantly impact the shape of the continuum and emission lines in the spectrum. We use the {\sc Python} package \texttt{WICKED} to remove the \textit{wiggle} from the spectrum, described in more detail in \citet{dumont25}. \texttt{WICKED} uses an aperture and annular integrated spectra templates, a power law, and a second-degree polynomial to model the spectrum. This model represents a \textit{wiggle} free model of the spectrum, which is used to create a \textit{wiggle} spectrum by subtracting this model from the spectrum. This wiggle spectrum is then fitted with a series of sinusoidal fits and then removed from the spectrum. Figure~\ref{fig:wicked} shows an example of \texttt{WICKED} for the removal of \textit{wiggle} from the brightest cube pixel spectrum of J0313-1806 (red).

We used a five-spaxel aperture spectrum template (black) and an eight-spaxel annular template (yellow), which are used in \texttt{WICKED} to create the best-fit model shown in blue. In the middle panel, the best-fit wiggle model is shown in red, and in the bottom panel, the corrected spectrum is shown in red. We ran \texttt{WICKED} with default parameters and a frequency prior of $8\, \mathrm{\mu m}^{-1}$ which leads to a better fit for the wiggle model. We set {\sc smooth\_spectrum} to "yes" during the flagging of spaxels with wiggles. This applies a $0.01 \, \mathrm{\mu m}$ mean smoothing to the spectrum, which improves the detection of wiggles in low signal-to-noise data.
We used a $3.5\sigma$ Fourier ratio in \texttt{WICKED}, leading to 103 spaxels in the datacube flagged as affected by wiggles. 

\begin{figure}
    \centering
    \includegraphics[width=0.95\linewidth]{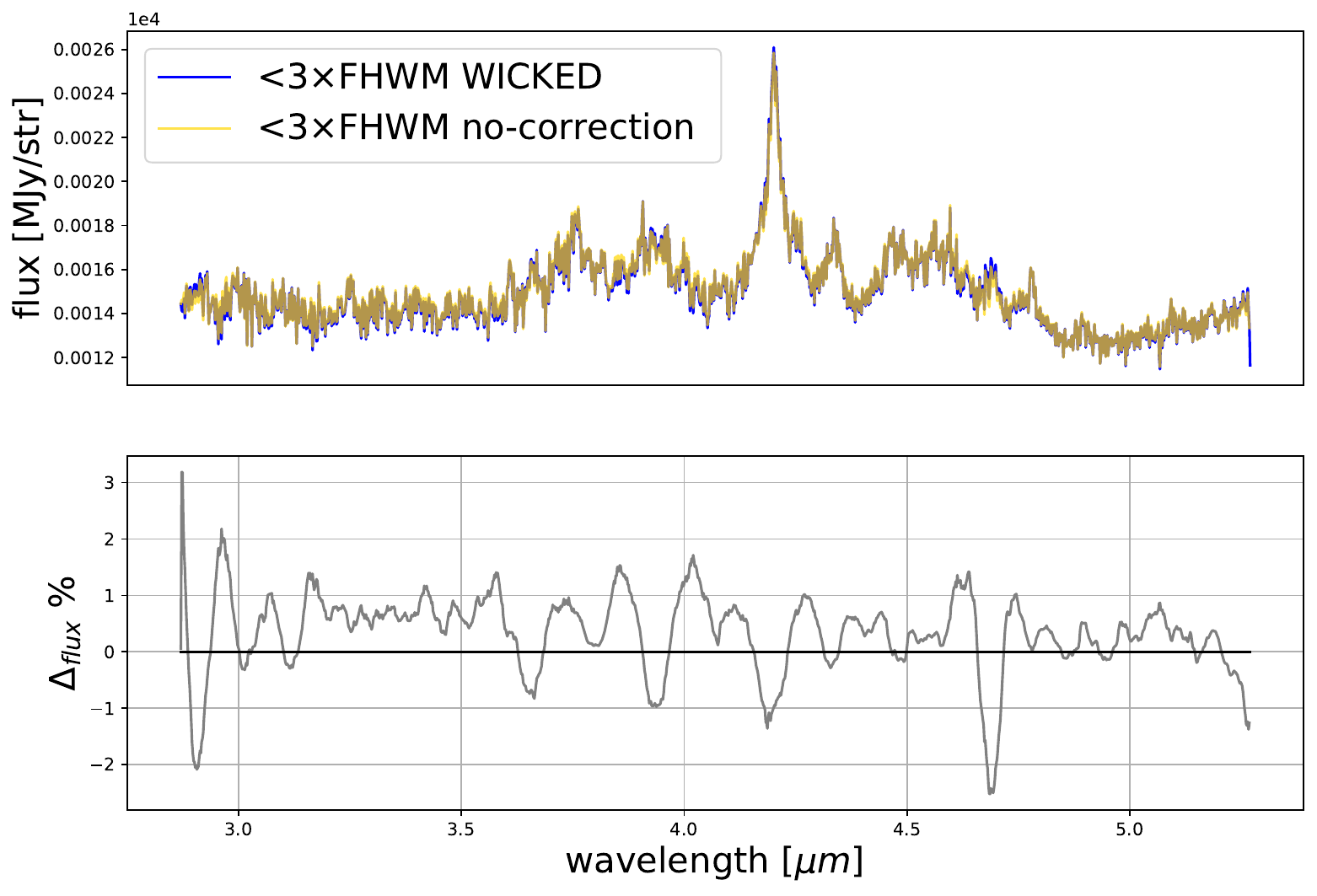}
    \caption{\textit{Top:} 3-spaxel aperture spectrum comparison for the datacube corrected with \texttt{WICKED} (blue) vs. uncorrected (yellow). \textit{Bottom:} Percentage difference (grey) between the uncorrected and \texttt{WICKED} data.
    }
\label{fig:wicked_vs_uncorrected}
\end{figure}

Figure~\ref{fig:wicked_vs_uncorrected} shows the comparison between a 3-spaxel ($\sim 3\times$FWHM) aperture extraction between the corrected with \texttt{WICKED} and the original datacube. We observe a mean difference of $\sim 0,3\%$ and a maximum difference of $\sim2\%$. This is within the difference $\leq5\%$ for the corrected vs. uncorrected data described in \citet{dumont25}.

\section{Determination of quasar spectrum extraction radius}
\label{sec:appendix_extract}

\begin{figure}
    \centering
    \includegraphics[width=0.95\linewidth]{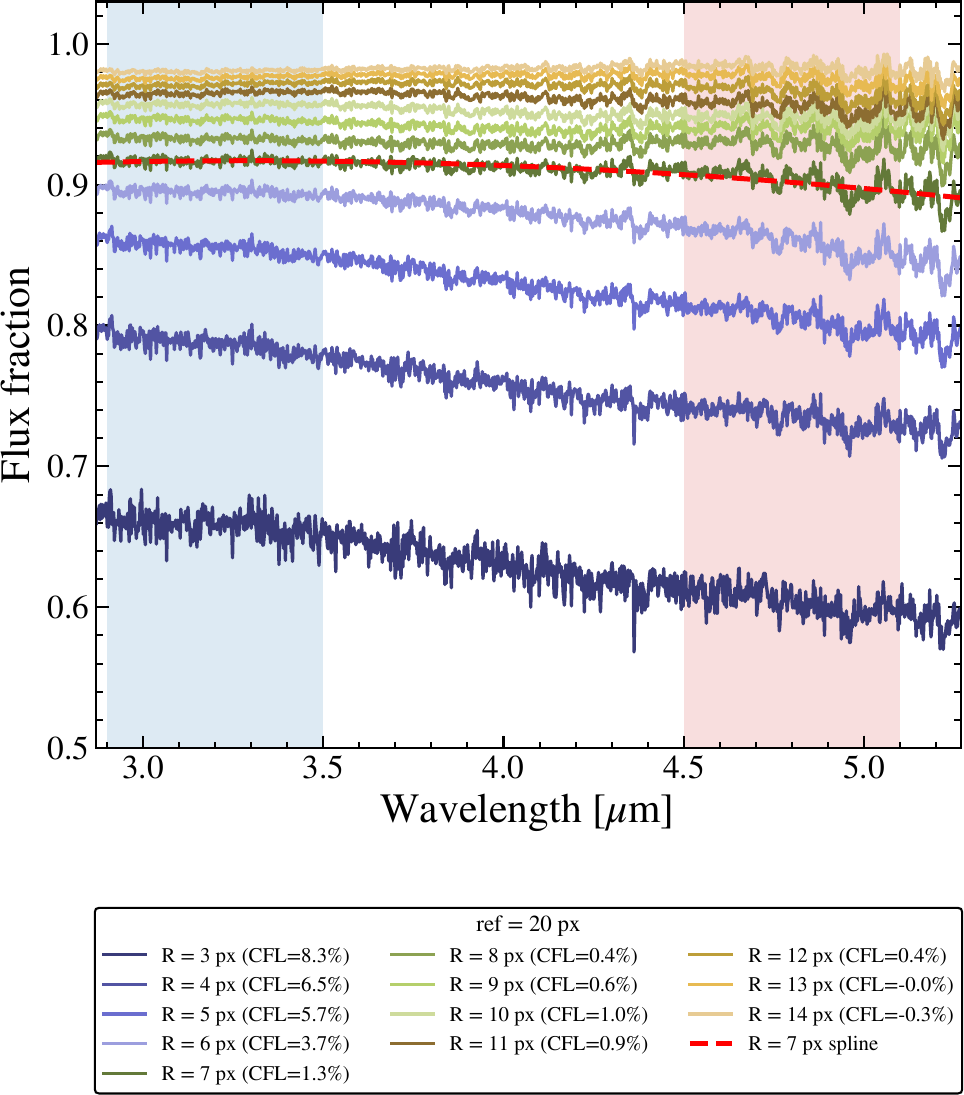}
\caption{Wavelength--dependent encircled--flux fractions for different aperture radii, shown relative to a large reference aperture of $1\farcs0$. The legend reports the corresponding chromatic flux losses as defined in the main text. A 7--pixel radius achieves percent-level chromatic flux loss while avoiding the increased background and extended--emission contamination present in larger apertures. The spline function fitted to the 7--pixel curve is shown as a red dashed line. The blue and red bands correspond to the regions we compute the chromatic flux losses from:} (2.9--3.5\,$\mu\mathrm{m}$) and (4.5--5.1\,$\mu\mathrm{m}$.) 

    \label{fig:apextract}
\end{figure}

In Fig.~\ref{fig:apextract} we show the wavelength--dependent flux fractions of the nuclear spectra extracted with apertures in the range 3--14 pixels, 0\farcs 15-- 0\farcs 70, relative to that obtained with a larger reference aperture of radius 20 pixels, i.e. 1\farcs 0. For smaller apertures, the downturn toward longer wavelengths reflects the fact that the PSF broadens with increasing wavelength, so a fixed aperture encloses a decreasing fraction of the total flux. Based on this figure, we adopted an extraction aperture with a radius of $0\farcs35$.

\section{Tentative detection of extended continuum emission}
\label{sec:appendix_cont}

\begin{figure}
    \centering
    \includegraphics[width=0.95\linewidth]{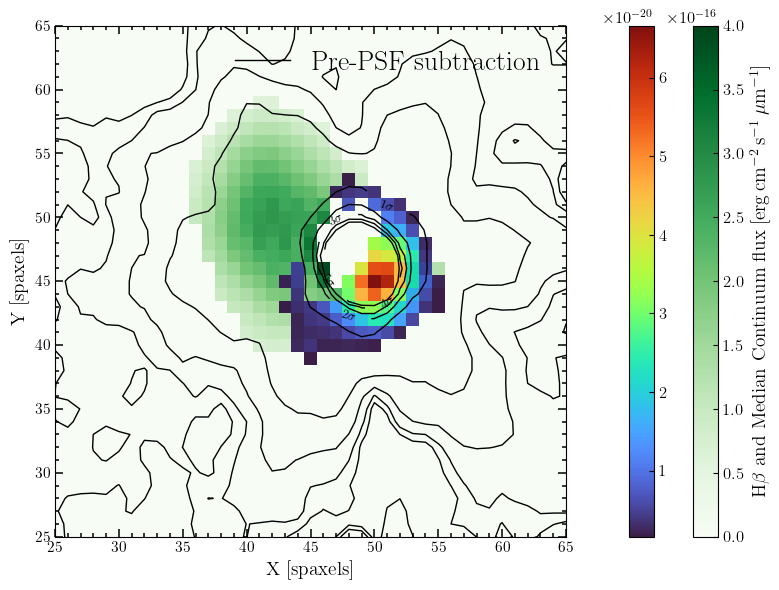}
    \caption{Median continuum in PSF-subtracted cube. The green coloured region corresponds to all islands of at least 5 contiguous spaxels with a mean continuum, measured over 3-4.2 $\mu m$, detected at the $1\sigma$ level. We compare it here to the contours of the initial non-PSF-subtracted quasar cube. We also indicate the location of the significant \hbetashort line emission enveloppe. The offset location of the continuum emission and its elliptical morphology indicate that it is likely not a PSF residual.}
    \label{fig:continuum}
\end{figure}

After the line and continuum fitting procedure, we searched for significant continuum emission: we searched for contiguous regions of at least 5 spaxels for which the mean continuum in the spectral window $\delta \lambda = 3-4.2 \, \mu m$ is at least detected at the 1$\sigma$ level. We confirm the detection of an elliptical nebula to the north-east of the quasar position. The relative positions of this low-surface brightness ellipse and the \hbetashort line emitting region are shown in Fig. \ref{fig:continuum}. We fitted a Sérsic profile to this continuum detected nebula. Its centroid is constrained to \texttt{x\_pix}$=44$ and \texttt{y\_pix}$=47$. We further find an ellipticity of $0.39\pm 0.01$, a Sérsic index of $n= 0.51 \pm 0.02$ and an effective radius of $r_e=7.05 \pm 0.10 \, \mathrm{pix}$. Assuming this structure is at the quasar redshift $z=7.6423$, the effective radius corresponds to a physical scale of $1.75 \pm 0.02$ kpc. The diffuse continuum emission could potentially trace a compact, moderately flattened companion of the quasar, with a shallow surface brightness profile typical of disturbed disks, tidal debris, irregular galaxies. This is further supported by the large offset between the quasar and the centre of the ellipsoid of $\sim 1.2$ kpc and the distinct morphologies of the \hbetashort line-emitting envelope and the smooth-profiled and symmetric detected continuum. The large dispersion in the \hbetashort shell further suggests velocity shear, a typical feature of interactions.

Given the absence of clear, distinctive spectral features to identify the redshift of this potential companion, we cannot exclude that it is located in the fore- or background. As most massive quasar hosts are expected to be compact and anchored in deep gravitational potentials, it is unlikely that the continuum corresponds to an irregular extension of the quasar's host galaxy. Furthermore, many high-redshift quasars observed with JWST/NIRSpec IFU so far appear to find themselves in some kind of interaction with a companion galaxy \citep[e.g,][]{marshall24,decarli24}.

\section{Spectra extracted from various emission regions in the NIRSpec IFU cube of J0313$-$1806}
\label{sec:appendix_specs}

\begin{figure}
    \centering
    \includegraphics[width=0.95\linewidth]{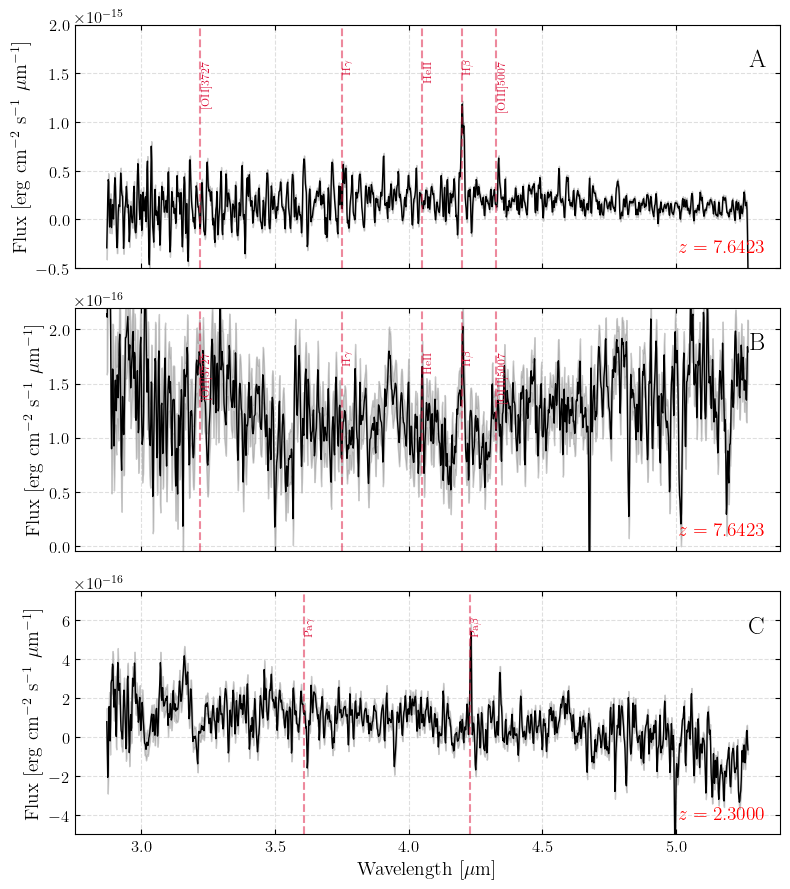}
    \caption{Spectra extracted from the PSF subtracted NIRSpec IFU cube within circular apertures of radii 0\farcs35 at the regions A, B and C (defined in Section \ref{sec:psf_model}). Region A corresponds to the \hbetashort nebula discussed in this paper. Region B corresponds to the extended continuum emission around to the north-east of the quasar. Region C corresponds to a foreground galaxy at $z=2.3$.}
    \label{fig:extract_spec}
\end{figure}

In Fig. \ref{fig:extract_spec} we display spectra extracted within 0\farcs35 of regions A, B and C defined in Section \ref{sec:psf_model}. 

\end{appendix}

\end{document}